\documentstyle[10pt,epsfig,dp_delphititle]{dp_delphi}
\makeindex
\def\DpPaperGroup{PH--EP}
\def\DpPaperRef{2009-014}
\def\DpDate{22 June 2009}
\def\DpAuthors{DELPHI Collaboration}
\def\DpSubmit{(Accepted by Eur. Phys. J. C)}
\def\DpTitle{Study of the Dependence   \\
of Direct Soft Photon Production \\
on the Jet Characteristics       \\
in Hadronic $Z^0$ Decays}
\def\DpComment{}
\def\DpEMail{}

\newcommand{\be}{\begin{equation}}
\newcommand{\ee}{\end{equation}}
\newcommand{\bd}{\begin{displaymath}}
\newcommand{\ed}{\end{displaymath}}
\newcommand{\bt}{\begin{tabular}}
\newcommand{\et}{\end{tabular}}
\newcommand{\efig}{\end{figure}}
\newcommand{\bc}{\begin{center}}
\newcommand{\ec}{\end{center}}

\begin{document}
%%%%%%%%%%%%%%%%%%%%%%%%%% They are a problem with Coll.Sty ?
\makeatletter
\makeatother
%%%%%%%%%%%%%%%%%%%%%%%%%% ??????????????????????????????????
%   Generate the title page

\begin{titlepage}
\pagenumbering{roman}

\CERNpreprint{\DpPaperGroup}{\DpPaperRef}   % Reference of the paper
\date{{\small\DpDate}}			    % Date of the paper
\title{\DpTitle}			    % Title of the paper
\address{\DpAuthors}			    % General name of the author(s)

\begin{shortabs}			    % Start the abstract
\noindent
%\input{abstract.tex}	% Paper abstract
%   abstract.tex
%
\noindent
%===================> Abstract     =====> To be filled <=====%
An analysis of the direct soft photon production rate as a function 
of the parent jet characteristics is presented, based on hadronic 
events collected by the DELPHI experiment at LEP1. The dependences of the 
photon rates on the jet kinematic characteristics (momentum, mass, etc.) 
and on the jet charged, neutral and total hadron multiplicities are reported. 
Up to a scale factor of about four, which characterizes the overall value 
of the soft photon excess, a similarity of 
the observed soft photon behaviour to that of the inner hadronic 
bremsstrahlung predictions is found for the momentum, mass, and 
jet charged multiplicity dependences. However for the dependence of the soft 
photon rate on the jet neutral and total hadron multiplicities a prominent 
difference is found for the observed soft photon signal as compared to the 
expected bremsstrahlung from final state hadrons. The observed linear increase 
of the soft photon production rate with the jet total hadron multiplicity and 
its strong dependence on the jet neutral multiplicity suggest that the rate 
is proportional to the number of quark pairs produced in the fragmentation 
process, with the neutral pairs being more effectively radiating than the
charged ones.
\end{shortabs}

\vfill

\begin{center}
\DpSubmit \ \\          % Horrible hack to allow to have DpSubmit empty
\DpComment \ \\
\DpEMail \ \\
\end{center}

\vfill
\clearpage

\headsep 10.0pt

\addtolength{\textheight}{10mm}
\addtolength{\footskip}{-5mm}

\begingroup
%			Commands to process the author names
%
\newcommand{\DpName}[2]{\hbox{#1$^{\ref{#2}}$},\hfill}
\newcommand{\DpNameTwo}[3]{\hbox{#1$^{\ref{#2},\ref{#3}}$},\hfill}
\newcommand{\DpNameThree}[4]{\hbox{#1$^{\ref{#2},\ref{#3},\ref{#4}}$},\hfill}
\newskip\Bigfill \Bigfill = 0pt plus 1000fill
\newcommand{\DpNameLast}[2]{\hbox{#1$^{\ref{#2}}$}\hspace{\Bigfill}}
%
%\small
\footnotesize
\noindent
\DpName{J.Abdallah}{LPNHE}
\DpName{P.Abreu}{LIP}
\DpName{W.Adam}{VIENNA}
\DpName{P.Adzic}{DEMOKRITOS}
\DpName{T.Albrecht}{KARLSRUHE}
\DpName{R.Alemany-Fernandez}{CERN}
\DpName{T.Allmendinger}{KARLSRUHE}
\DpName{P.P.Allport}{LIVERPOOL}
\DpName{U.Amaldi}{MILANO2}
\DpName{N.Amapane}{TORINO}
% ...
\normalsize
\endgroup
\newpage
\titlefoot{Department of Physics and Astronomy, Iowa State
     University, Ames IA 50011-3160, USA
    \label{AMES}}
\titlefoot{Physics Department, Universiteit Antwerpen,
     Universiteitsplein 1, B-2610 Antwerpen, Belgium
    \label{ANTWERP}}
\titlefoot{IIHE, ULB-VUB,
     Pleinlaan 2, B-1050 Brussels, Belgium
    \label{BRUSSELS}}
\titlefoot{Physics Laboratory, University of Athens, Solonos Str.
     104, GR-10680 Athens, Greece
    \label{ATHENS}}
\titlefoot{Department of Physics, University of Bergen,
     All\'egaten 55, NO-5007 Bergen, Norway
    \label{BERGEN}}
\titlefoot{Dipartimento di Fisica, Universit\`a di Bologna and INFN,
     Viale C. Berti Pichat 6/2, IT-40127 Bologna, Italy
    \label{BOLOGNA}}
\titlefoot{Centro Brasileiro de Pesquisas F\'{\i}sicas, rua Xavier Sigaud 150,
     BR-22290 Rio de Janeiro, Brazil
    \label{BRASIL-CBPF}}
\titlefoot{Inst. de F\'{\i}sica, Univ. Estadual do Rio de Janeiro,
     rua S\~{a}o Francisco Xavier 524, Rio de Janeiro, Brazil
    \label{BRASIL-IFUERJ}}
\titlefoot{Coll\`ege de France, Lab. de Physique Corpusculaire, IN2P3-CNRS,
     FR-75231 Paris Cedex 05, France
    \label{CDF}}
\titlefoot{CERN, CH-1211 Geneva 23, Switzerland
    \label{CERN}}
\titlefoot{Institut Pluridisciplinaire Hubert Curien, Universit\'e de Strasbourg,
     FR-67037 Strasbourg Cedex 2, France
    \label{CRN}}
\titlefoot{Now at DESY-Zeuthen, Platanenallee 6, D-15735 Zeuthen, Germany
    \label{DESY}}
\titlefoot{Institute of Nuclear Physics, N.C.S.R. Demokritos,
     P.O. Box 60228, GR-15310 Athens, Greece
    \label{DEMOKRITOS}}
\titlefoot{FZU, Inst. of Phys. of the C.A.S. High Energy Physics Division,
     Na Slovance 2, CZ-182 21, Praha 8, Czech Republic
    \label{FZU}}
\titlefoot{Dipartimento di Fisica, Universit\`a di Genova and INFN,
     Via Dodecaneso 33, IT-16146 Genova, Italy
    \label{GENOVA}}
\titlefoot{Institut des Sciences Nucl\'eaires, IN2P3-CNRS, Universit\'e
     de Grenoble 1, FR-38026 Grenoble Cedex, France
    \label{GRENOBLE}}
\titlefoot{Helsinki Institute of Physics and Department of Physical Sciences,
     P.O. Box 64, FIN-00014 University of Helsinki, 
     \indent~~Finland
    \label{HELSINKI}}
\titlefoot{Joint Institute for Nuclear Research, Dubna, Head Post
     Office, P.O. Box 79, RU-101 000 Moscow, Russian Federation
    \label{JINR}}
\titlefoot{Institut f\"ur Experimentelle Kernphysik,
     Universit\"at Karlsruhe, Postfach 6980, DE-76128 Karlsruhe,
     Germany
    \label{KARLSRUHE}}
\titlefoot{Institute of Nuclear Physics PAN,Ul. Radzikowskiego 152,
     PL-31142 Krakow, Poland
    \label{KRAKOW1}}
\titlefoot{Faculty of Physics and Nuclear Techniques, University of Mining
     and Metallurgy, PL-30055 Krakow, Poland
    \label{KRAKOW2}}
%\titlefoot{Universit\'e de Paris-Sud, Lab. de l'Acc\'el\'erateur
%    Lin\'eaire, IN2P3-CNRS, B\^{a}t. 200, FR-91405 Orsay Cedex, France
\titlefoot{LAL, Univ Paris-Sud, CNRS/IN2P3, Orsay, France
    \label{LAL}}
\titlefoot{School of Physics and Chemistry, University of Lancaster,
     Lancaster LA1 4YB, UK
    \label{LANCASTER}}
\titlefoot{LIP, IST, FCUL - Av. Elias Garcia, 14-$1^{o}$,
     PT-1000 Lisboa Codex, Portugal
    \label{LIP}}
\titlefoot{Department of Physics, University of Liverpool, P.O.
     Box 147, Liverpool L69 3BX, UK
    \label{LIVERPOOL}}
\titlefoot{Dept. of Physics and Astronomy, Kelvin Building,
     University of Glasgow, Glasgow G12 8QQ, UK
    \label{GLASGOW}}
\titlefoot{LPNHE, IN2P3-CNRS, Univ.~Paris VI et VII, Tour 33 (RdC),
     4 place Jussieu, FR-75252 Paris Cedex 05, France
    \label{LPNHE}}
\titlefoot{Department of Physics, University of Lund,
     S\"olvegatan 14, SE-223 63 Lund, Sweden
    \label{LUND}}
\titlefoot{Universit\'e Claude Bernard de Lyon, IPNL, IN2P3-CNRS,
     FR-69622 Villeurbanne Cedex, France
    \label{LYON}}
\titlefoot{Dipartimento di Fisica, Universit\`a di Milano and INFN-MILANO,
     Via Celoria 16, IT-20133 Milan, Italy
    \label{MILANO}}
\titlefoot{Dipartimento di Fisica, Univ. di Milano-Bicocca and
     INFN-MILANO, Piazza della Scienza 3, IT-20126 Milan, Italy
    \label{MILANO2}}
\titlefoot{IPNP of MFF, Charles Univ., Areal MFF,
     V Holesovickach 2, CZ-180 00, Praha 8, Czech Republic
    \label{NC}}
\titlefoot{NIKHEF, Postbus 41882, NL-1009 DB
     Amsterdam, The Netherlands
    \label{NIKHEF}}
\titlefoot{National Technical University, Physics Department,
     Zografou Campus, GR-15773 Athens, Greece
    \label{NTU-ATHENS}}
\titlefoot{Physics Department, University of Oslo, Blindern,
     NO-0316 Oslo, Norway
    \label{OSLO}}
\titlefoot{Dpto. Fisica, Univ. Oviedo, Avda. Calvo Sotelo
     s/n, ES-33007 Oviedo, Spain
    \label{OVIEDO}}
\titlefoot{Department of Physics, University of Oxford,
     Keble Road, Oxford OX1 3RH, UK
    \label{OXFORD}}
\titlefoot{Dipartimento di Fisica, Universit\`a di Padova and
     INFN, Via Marzolo 8, IT-35131 Padua, Italy
    \label{PADOVA}}
\titlefoot{Rutherford Appleton Laboratory, Chilton, Didcot
     OX11 OQX, UK
    \label{RAL}}
\titlefoot{Dipartimento di Fisica, Universit\`a di Roma II and
     INFN, Tor Vergata, IT-00173 Rome, Italy
    \label{ROMA2}}
\titlefoot{Dipartimento di Fisica, Universit\`a di Roma III and
     INFN, Via della Vasca Navale 84, IT-00146 Rome, Italy
    \label{ROMA3}}
\titlefoot{DAPNIA/Service de Physique des Particules,
     CEA-Saclay, FR-91191 Gif-sur-Yvette Cedex, France
    \label{SACLAY}}
\titlefoot{Instituto de Fisica de Cantabria (CSIC-UC), Avda.
     los Castros s/n, ES-39006 Santander, Spain
    \label{SANTANDER}}
\titlefoot{Inst. for High Energy Physics, Serpukov
     P.O. Box 35, Protvino, (Moscow Region), Russian Federation
    \label{SERPUKHOV}}
\titlefoot{J. Stefan Institute, Jamova 39, SI-1000 Ljubljana, Slovenia
    \label{SLOVENIJA1}}
\titlefoot{Laboratory for Astroparticle Physics,
     University of Nova Gorica, Kostanjeviska 16a, SI-5000 Nova Gorica, Slovenia
    \label{SLOVENIJA2}}
\titlefoot{Department of Physics, University of Ljubljana,
     SI-1000 Ljubljana, Slovenia
    \label{SLOVENIJA3}}
\titlefoot{Fysikum, Stockholm University,
     Box 6730, SE-113 85 Stockholm, Sweden
    \label{STOCKHOLM}}
\titlefoot{Dipartimento di Fisica Sperimentale, Universit\`a di
     Torino and INFN, Via P. Giuria 1, IT-10125 Turin, Italy
    \label{TORINO}}
\titlefoot{INFN,Sezione di Torino and Dipartimento di Fisica Teorica,
     Universit\`a di Torino, Via Giuria 1,
     IT-10125 Turin, Italy
    \label{TORINOTH}}
\titlefoot{Dipartimento di Fisica, Universit\`a di Trieste and
     INFN, Via A. Valerio 2, IT-34127 Trieste, Italy
    \label{TRIESTE}}
\titlefoot{Istituto di Fisica, Universit\`a di Udine and INFN,
     IT-33100 Udine, Italy
    \label{UDINE}}
\titlefoot{Univ. Federal do Rio de Janeiro, C.P. 68528
     Cidade Univ., Ilha do Fund\~ao
     BR-21945-970 Rio de Janeiro, Brazil
    \label{UFRJ}}
\titlefoot{Department of Radiation Sciences, University of
     Uppsala, P.O. Box 535, SE-751 21 Uppsala, Sweden
    \label{UPPSALA}}
\titlefoot{IFIC, Valencia-CSIC, and D.F.A.M.N., U. de Valencia,
     Avda. Dr. Moliner 50, ES-46100 Burjassot (Valencia), Spain
    \label{VALENCIA}}
\titlefoot{On leave of absence from ITEP, 117259 Moscow, Russian Federation
    \label{ITEP}}
\titlefoot{Institut f\"ur Hochenergiephysik, \"Osterr. Akad.
     d. Wissensch., Nikolsdorfergasse 18, AT-1050 Vienna, Austria
    \label{VIENNA}}
\titlefoot{Inst. Nuclear Studies and University of Warsaw, Ul.
     Hoza 69, PL-00681 Warsaw, Poland
    \label{WARSZAWA}}
\titlefoot{Now at University of Warwick, Coventry CV4 7AL, UK
    \label{WARWICK}}
\titlefoot{Fachbereich Physik, University of Wuppertal, Postfach
     100 127, DE-42097 Wuppertal, Germany \\
\noindent
{$^\dagger$~deceased}
    \label{WUPPERTAL}}
\addtolength{\textheight}{-10mm}
\addtolength{\footskip}{5mm}
\clearpage

\headsep 30.0pt
\end{titlepage}

%%%%%%%%%%%%%%%%%%%%%%%%%
%
%	Change for the document body
%\pagestyle{heading}				    % for page numbering
\pagenumbering{arabic}				    % page numbering in number
\setcounter{footnote}{0}			    %
\large
%\linenumbers
%\input{document.tex}	% The body of the document.
%   document.tex
%
%*****************************************************************************
\section{Introduction}
Recent analysis of the soft photon production in hadronic decays of the $Z^0$
studied with the DELPHI detector at LEP1 \cite{aspdel} revealed 
a significant excess of soft photons deep inside jets as compared to the 
predictions of parton shower models 
\cite{jetset,jetset2,jetset3,ariadne,herwig} for the photon rates 
induced by hadrons decaying radiatively (most of the photons
coming from $\pi^0$'s). The photon kinematic range was defined in 
\cite{aspdel} as: $0.2 < E_{\gamma} < 1$ GeV, $p_T < 80$ MeV/$c$, the $p_T$ 
being the photon transverse momentum with respect to the parent jet direction. 
Furthermore, the observed signal was much greater than the level 
of the inner hadronic bremsstrahlung, which according to the QED 
predictions (see \cite{low,gribov}) was expected to be the dominant 
source of the direct soft photons in this kinematic region. Expressed in terms 
of the predicted bremsstrahlung rate, the observed signal was found to be
$3.4 \pm 0.2 \pm 0.8$ for the data uncorrected for the detection efficiency,
and $4.0 \pm 0.3 \pm 1.0$ for the corrected data (the first errors are 
statistical, the second ones are systematic). 

The observation of the excess of soft photons in hadronic events 
of $Z^0$ decays reported in \cite{aspdel} is indeed a further contribution 
to the collection of the anomalous soft photon effects found earlier
in reactions of multiple hadron production in several hadronic beam 
experiments at high energy \cite{wa27,na22,wa83,wa91,wa91a,wa102}, 
all at the photon 
c.m.s. rapidities $ y >1.2 $ \footnote{For the sake of completeness we mention 
two studies of soft photons at central and slightly backward c.m.s. rapidities 
\cite{bnl,helios} in which no photon excess over the expected bremsstrahlung 
level was observed.}. Known for more than 20 years, they however 
still lack a theoretical explanation, in spite of being under active 
investigation. Reviews of the theoretical approaches to the problem can be 
found in \cite{pis,lich} (see also the references [13-33] in \cite{aspdel}).

On the other hand, no deviation of the photon production rates and/or other 
radiation characteristics has been observed compared to the predictions 
based on QED for effects of pure electroweak nature. For 
example, the electron inner bremsstrahlung in $e^+ e^-$ collisions at LEP 
(initial state radiation, ISR) was an important experimental effect, with
which all the LEP experiments had to contend. No deviation of the ISR
characteristics from those expected from theory was observed, either at the
$Z^0$ or at high energy (see e.g. the DELPHI studies \cite{isr,isr2,isr3}). 
A direct study of the muon inner bremsstrahlung 
in $\mu^+ \mu^-$ decays of the $Z^0$ 
(final state radiation, FSR) in events collected by the DELPHI experiment at 
LEP1, with the same experimental method as employed in \cite{aspdel}, has shown 
a good agreement of the observed photon production characteristics with 
those expected from the bremsstrahlung predictions \cite{muonbrems}. 

Thus, the soft photon anomaly seems to be restricted to the processes of 
multiple hadron production, i.e. it is rooted in strong interaction physics. 
Nevertheless it is clear that the development of a theory resolving this 
long-standing problem currently requires further experimental information. 
The process of $e^+ e^-$ annihilation to hadrons, in which events  
with well defined jet structure are produced, presents 
a suitable opportunity to meet the challenge. Therefore this paper continues
the investigation of events from reaction 
\begin{equation}
e^+e^- \rightarrow Z^0 \rightarrow \rm{direct~ soft ~\gamma + hadrons}
\end{equation}
begun in paper \cite{aspdel}. 
The aim of the present analysis is the study of several dependences 
of the direct soft photon production on the parent jet characteristics,
to which various models treating the anomalous radiation may be sensitive.
Among these characteristics are the jet momentum, mass, net charge and 
the jet particle multiplicities. The last dependence is presented 
subdivided into three branches: dependences on charged, neutral and total 
(neutral + charged) multiplicities of the jet. The kinematic
region of this analysis is kept the same as in \cite{aspdel}. Based on the 
results of this study, indications for a possible localization of the source 
of anomalous soft photons are obtained: such a source may be the creation
out of the QCD vacuum of the lightest ($u, d$) quarks, with their further 
evolution during the processes of the parton shower development and
hadronization.

This paper is organized as follows. Section 2 deals with the calculation of 
the inner hadronic bremsstrahlung. Section 3 provides a description of
the apparatus, software, and the experimental method applied. Section 4
describes the selection cuts and data samples. In section 5 the definition
of variables used in the analysis is given. Systematic uncertainties arising 
from various elements of the analysis method, and their estimates are 
presented in section 6. In section 7 the main results of the analysis are 
given.  Discussion of the obtained results and their possible interpretation, 
with emphasis on the strong signal dependence on the jet neutral multiplicity, 
are given in section 8. Finally, section~9 provides a summary and conclusions.

\section{ Bremsstrahlung calculations}
The production rate for both ISR and inner bremsstrahlung from final
hadronic states in the soft photon region can be calculated simultaneously 
using a universal formula derived from Low \cite{low} with a modification 
suggested by Haissinski \cite{hais,grab,tim}:
\begin{equation}
\frac{dN_{\gamma}}{d^{3}\vec{k}}
=
\frac{\alpha}{(2 \pi)^2} \frac{1}{E_\gamma}
\int d^3 \vec{p}_{1} . . . d^3 \vec{p}_{N}
\sum_{i,j} \eta_{i} \eta_{j}
\frac{(\vec{p}_{i \bot} \cdot \vec{p}_{j \bot}) }{ ( P_{i} K )  ( P_{j} K )}
\frac{ d N_{hadrons}}{ d^{3} \vec{p}_{1} ... d^{3} \vec{p}_{N}} ~.
\end{equation}
Here $K$ and $\vec{k}$ denote photon four- and
three-momenta, $P$ are the four-momenta of beam $e^+, e^-$ and $N$
charged outgoing hadrons, and $\vec{p}_1$ ...$\vec{p}_N$ are the
three-momenta of the hadrons;
$\vec{p}_{i \bot} = \vec{p}_i-(\vec{n} \cdot \vec{p}_i) \cdot \vec{n}$ and
~$\vec{n}$ is the photon unit vector, $\vec{n} = \vec{k}/k $;
$\eta=1$ for the beam electron and for positive outgoing hadrons,
$\eta=-1$ for the beam positron and negative outgoing hadrons,
and the sum is extended over all the $N+2$ charged particles involved;
the last factor in the integrand is a differential hadron production rate.

Note, formula (2) is completely equivalent, from a theoretical point of view,
to the analogous one, applied for calculation of the inner hadronic 
bremsstrahlung in \cite{wa27,na22,pis} and considered to be standard in 
textbooks on electrodynamics. It differs from the latter by the numerator 
$(\vec{p}_{i \bot} \cdot \vec{p}_{j \bot})$, 
used to replace the four-dimensional scalar product $-( P_{i} P_{j} )$. 
When dealing with relativistic particles, the advantage of such a 
replacement is essential and is based on the following. 
Both formulae operate, in general, by terms of big absolute values adjusted 
in such a way that they cancel each other in the sum almost completely due to
$\eta_i, \eta_j$ alternate signs. However this ``fine tuning" 
which is rooted in the gauge invariance of electrodynamics and
reflects the charge conservation law, is achieved in numerical calculations 
more easily with formula (2).
Therefore, when using detected particle spectra in the bremsstrahlung 
calculations, formula (2) is more stable with respect to the particle loss
and measurement errors as compared to the standard one. 
Moreover, even in the case of using 
precise Monte Carlo spectra for the calculations, as was done in 
\cite{aspdel,wa83,wa91,wa91a,wa102,muonbrems}, the implementation of 
formula (2) should be preferred in computing the bremsstrahlung as giving 
smaller fluctuations of the sum terms for the specific particle and photon 
momentum configurations leading to extremely low values of denominators 
in formula (2) (so called collinear singularity).

Formula (2), as well as its standard analog, describes both initial 
state radiation from the colliding $e^+ e^-$, and the inner bremsstrahlung 
from the final hadronic states. However, it was demonstrated in \cite{aspdel} 
that the ISR is rather small in the range of the photon polar angles 
to the beam $\Theta_{\gamma}$ used in this analysis (barrel region), 
being about 1.5\% of the total hadronic inner bremsstrahlung. The situation 
changes little even when there are very few charged particles inside the jet. 
For example, for charged jet multiplicities between 0 and 2 (which corresponds 
to the first bin of the photon rate distribution over the $N_{ch}$ variable
defined below, see Sect. 5.3) the ISR rate in the chosen kinematic range is 
at the level of about 4\% of the inner hadronic bremsstrahlung yield in 
this bin. Therefore the ISR contribution is marginal in the predicted 
bremsstrahlung rates.

Similarly, the yield of final state radiation from quarks of $Z^0$ 
disintegrations, calculated within the standard perturbative approach
implemented in the LUND fragmentation model 
\cite{jetset,jetset2,jetset3,lundgam}, is small too. 
It was shown in \cite{aspdel} where this approach was used
to evaluate the bremsstrahlung radiation off quarks, that it is at the level 
of 3\% of the inner hadronic bremsstrahlung within the kinematic range
considered. The reasons for this suppression are the fractional quark charges 
(which give an attenuation factor of about 1/4)
and large quark virtualities which are intrinsic for this approach.
                            
The treatment of the three listed bremsstrahlung sources (inner hadronic 
bremsstrahlung, ISR and the radiation off quarks of $Z^0$ disintegrations)
was different in the Monte Carlo (MC) stream  as described below (Sect. 3.2).
                                                    
\section{ Experimental technique}
\subsection{The DELPHI detector}
The DELPHI detector is described in detail in \cite{delphi1,delphi2}. 
The following is a brief description of the subdetector units relevant for 
this analysis: the main tracker of the DELPHI detector, the Time Projection 
Chamber (TPC), the barrel electromagnetic calorimeter, the High density
Projection Chamber (HPC), and the hadronic calorimeter (HCAL) 
 
In the DELPHI reference frame the $z$ axis is taken
along the direction of the $e^-$ beam. The angle $\Theta$ is the polar
angle defined with respect to the $z$-axis, $\Phi$ is the azimuthal angle
around this axis and $R$ is the distance from this axis.

The TPC covered the angular 
range from $20^{\circ}$ to $160^{\circ}$ in $\Theta$ and extended
from 30 cm to 122 cm in $R$. It provided up to 16 space points for
pattern recognition and ionization information extracted from 192 wires. The 
momentum threshold for charged particles entering the TPC was approximately 
100 MeV/$c$.

The HPC was used for the detection of high energy photons, which originate
in hadronic events mainly from the decays of neutral pions. The HPC lay 
immediately outside the tracking detectors and covered the angles 
$\Theta$ from $43^{\circ}$ to $137^{\circ}$.
It had eighteen radiation lengths for perpendicular incidence, and
its energy resolution was $\Delta E/E = 0.31/E^{0.44}\oplus 0.027$ where
$E$ is in units of GeV \cite{pi0}. It had a high granularity and provided a
sampling of shower energies from nine layers in depth. The angular precisions
for high energy photons were $\pm 1.0$ mrad in $\Theta$ and $\pm1.7$
mrad in $\Phi$.

The HCAL was installed in the return yoke of the DELPHI solenoid and provided
a relative precision on the measured energy of 
$\Delta E/E = 1.12/\sqrt{E} \oplus 0.21$. It was used for the detection 
of $K^0_L$'s and neutrons. 

\subsection{ Monte Carlo generators}
The principal Monte Carlo data sets used in this analysis were produced
with the JETSET 7.3 PS generator \cite{jetset,jetset2,jetset3}, based on 
the LUND string model \cite{lundstring}, with parameters adjusted
according to previous QCD studies \cite{tun1,tun1a,tun2}. 
For the test of possible systematic biases, another standard generator, 
ARIADNE 4.6 \cite{ariadne} with parameters adjusted by the DELPHI tuning 
\cite{tun2} was also used \footnote{As noticed in \cite{aspdel}, ARIADNE tends 
to underestimate the production of photons in the range of the photon 
$p_T < 80$~MeV/$c$. A special test, which exploited the $SU(2)$ symmetry of 
the strong interactions, using artificial photons produced from charged pions 
(similar to that described in Sect. 6.5 of \cite{aspdel}), has shown better
performance of JETSET versus variables under study, as compared to ARIADNE.
This explains why JETSET was chosen as the principal generator in this 
analysis.}.

No generation of bremsstrahlung photons from the final state hadrons was
implemented in these MC generators. On the other hand,  
initial state radiation and photon radiation from quarks of $Z^0$ 
disintegrations calculated with the photon implementation in JETSET 
\cite{lundgam} were involved in all the generations.

The generated events were fed into the DELPHI detector simulation program
DELSIM \cite{delphi2} in order to produce data which are as close as possible
to the real raw data. These data were then treated by the reconstruction and
analysis programs in exactly the same way as the real data.

In order to evaluate the contamination from the $Z^0 \rightarrow \tau^+ \tau^-$
channel MC events produced with the KORALZ 4.0 generator \cite{koralz} 
and passed through a full detector simulation and the analysis procedure
were used. 

Finally, the generator DYMU3 \cite{dymu3,dymu3a} was used to check the
applicability of formula (2) in our kinematic region, see Sect. 6.3.

\subsection{Detection of soft photons}
The experimental technique employed for the detection of soft photons
in this analysis was the same as in \cite{aspdel}, based on the reconstruction 
of the photons converted in front of the TPC. 
The following is a brief description of the method implemented for the 
photon reconstruction and main characteristics of the detected photons
obtained with it.

The photon conversions were reconstructed
by an algorithm that examined tracks reconstructed in the TPC. A search was 
made along each TPC track for the point where the tangent of its trajectory 
points directly to the beam spot in the $R\Phi$ projection.
Under the assumption that the opening angle of the electron-positron pair
is zero, this point represented a possible photon conversion position at
radius $R$. All tracks which had a solution $R$ that was more than one
standard deviation away from the primary vertex, as defined by the beam spot,
were considered to be conversion candidates. If two oppositely charged
conversion candidates were found with compatible conversion point
parameters they were linked together to form the converted photon. The
following selection criteria were imposed:
\begin{itemize}
 \item the $\Phi$ difference between the two conversion points was at most
30 mrad;
 \item the difference between the polar angles $\Theta$ of the two tracks
was at most 15 mrad;
 \item at least one of the tracks should have no associated hits in front
of the reconstructed mean conversion radius.
\end{itemize}
For the pairs fulfilling these criteria a $\chi^2$ was calculated from
$\Delta \Theta, \Delta \Phi$ and the difference of the reconstructed
conversion radii $\Delta R$ in order
to find the best combinations in cases where there were ambiguous
associations. A constrained fit was then applied to the electron-positron
pair candidate which forced a common conversion point with zero opening
angle and collinearity between the momentum sum and the line from the
beam spot to the conversion point.

The quality of the photon reconstruction in both, the real data (RD)
and MC events, the latter being produced as described in Sect. 3.2, 
was high, as can be judged comparing $\pi^0$ peaks in the RD and MC 
$\gamma - \gamma$ mass distributions obtained with converted photons
and shown in Fig. 8 of \cite{aspdel}. The almost precise identity 
of these peaks, together with their widths of less than 5 MeV/$c^2$, 
demonstrate that the detection and analysis procedures of the converted 
photons in the DELPHI detector are well understood; 
this statement is supported also by the results of the DELPHI papers 
\cite{pi0,bstar}, in which the converted photons were involved in the analysis.
   
Selection of photons for this analysis was done under the following cuts:
\begin{itemize}
 \item 20$^{\circ} \leq \Theta_{\gamma} \leq 160^{\circ}$;
 \item 5 cm $\leq R_{conv} \leq 50$ cm, where $R_{conv}$ is the conversion
radius;
 \item 200 MeV $\leq E_{\gamma} \leq 1$ GeV.
\end{itemize} 

The photon detection efficiency, i.e. conversion probability combined with the 
reconstruction efficiency, was determined with the MC events and parameterized 
against two variables. The first variable, $E_{\gamma}$, was used to describe
a fast variation of the efficiency within the energy range under study, from 
almost zero at 0.2 GeV up to 5 - 6\% at 1~GeV (a typical behaviour of the 
efficiency with $E_{\gamma}$ can be seen in Fig. 1 of paper \cite{aspdel}). 
For interpolation of the efficiency,
it was fitted by a 2nd order polynomial or by the form
$ a - b \times \exp[ -c(E_\gamma-0.2) ] $ with a $\chi^2$/n.d.f. close to 1 
in both cases; the difference in the corrected photon rates 
obtained with the two interpolation curves was about 2\%.

The second variable of the efficiency parameterization is related to the jet 
characteristic under investigation, i.e. the efficiencies were determined 
separately in every bin of the jet parameter under study. The weak dependences 
of the photon detection efficiency on several additional 
variables treated in \cite{aspdel} (the photon polar angle to the beam, 
$\Theta_{\gamma}$, the photon polar angle to the parent jet axis, 
$\theta_{\gamma}$, etc.) were decided to be averaged over in this analysis.

The validity of the efficiency finding can be considered as confirmed
by the results of DELPHI paper \cite{muonbrems} in which the inner
bremsstrahlung off muons from $Z^0$ dimuon decays was studied by
applying the efficiencies obtained as described above, and  the photon signal 
was found to be in good agreement with the theoretical expectations.
More generally, the muon inner bremsstrahlung study \cite{muonbrems}, 
being carried out with the same methods of photon detection and analysis 
as in the current study, gives them further credibility.

The accuracy of the converted photon energy measurement was about
$\pm$1.6\% in the given kinematic range, as follows from the MC studies. 
This estimate was confirmed by extracting the photon energy resolution
from the widths of $\pi^0$ peaks in the RD and MC $\gamma - \gamma$ mass
distributions obtained with converted photons as shown in Fig. 8 of 
\cite{aspdel}.
 
The angular precision of the photon direction reconstruction was determined 
with the MC data and was found to be of a Breit-Wigner shape, as expected 
for the superposition of many Gaussian distributions of varying width 
\cite{eadie}. The full widths ($\Gamma$'s) of the $\Delta \Theta_{\gamma}$ and
$\Delta \Phi_{\gamma}$ distributions were 4 and 5 mrad, respectively
(for details see \cite{aspdel}).

\section{Data selection} 
Events involving the hadronic decays of the $Z^0$ from the DELPHI data
of the 1992 to 1995 running periods were used in this analysis.
                                                                                
Selection of the hadronic events was standard, based on large event
charged multiplicity ($N_{ch}^{evt} \geq 5$) and high visible energy
($E_{vis} \geq 0.2 E_{cm}$).
In addition, the condition
$|\cos \Theta_{thrust}| < 0.95$ was imposed, where
$ \Theta_{thrust}$ is the angle between the thrust axis and the beam direction.
These criteria correspond to an efficiency of $(95.2 \pm 0.2) \%$.
with a general $Z^0 \rightarrow \tau^+ \tau^-$ contamination of 
$(0.5 \pm 0.1)\%$. The apparently low $\tau$ background is concentrated
in small multiplicity events and may contaminate essentially the low
jet multiplicity samples. Therefore a further suppression of the
$\tau$ events has been achieved as described below. 

First, the electron and muon anti-tagging was applied to the events of 
low multiplicity, defined as having $N_{ch}^{evt} \leq 7$. This method 
decreased the $\tau \tau$ background by 25\%. Then two additional selections 
aimed at the $\tau \tau$ events suppression were implemented. 
The first one required that the jet masses in the low multiplicity events 
exceeded 2 GeV/$c^2$, with an exception for the jets having two or less charged
particles: for such jets the lower mass cut was weakened, being at 1 GeV/$c^2$.
The second $\tau \tau$ rejection method was 
based on the impact parameter analysis using the fact
that the fraction of $\tau \tau$ events with $N_{ch}^{evt} \geq 5$ 
(i.e. surpassing the minimal $N_{ch}^{evt}$ multiplicity cut described above) 
is dominated by the $\tau$ hadronic decays containing at least one secondary
interaction of the decay products which increases the originally low
$\tau \tau$ event multiplicity. This leads to a considerably increased 
value of the sum of the track impact parameters in the $R\Phi$ projection in 
the $\tau \tau$ events as compared to hadronic events of 
the same multiplicity. Thus, only those small multiplicity events were 
selected in which the sum of the impact parameters in the $R\Phi$ projection 
over all charged particle tracks was below 1 cm (an analogous cut 
on the sum of the $z$ projection impact parameters was found to be ineffective 
for further $\tau \tau$ background suppression).
Together with the electron and muon anti-tagging these selections
resulted in additional $\tau \tau$ background suppression 
by a factor of 5.9, while keeping the hadronic event efficiency 
at the level of $(94.7 \pm 0.2) \%$. The differential rates of photons
from the $\tau \tau$ background (in the bins of the jet multiplicity variables 
defined in Sec. 5) will be given below, in the analysis section. It should be
noted, that being deduced from $\tau \tau$ MC events they can be 
underestimated (by an unknown factor, not exceeding however 1.1-1.6
as follows from our special study of $\tau \tau$ events), if  
anomalous soft photons are produced in the hadronic tau decays also. 

A total of 3~435~173 events of real data was selected under these cuts
and compared to $12.1\times 10^6$ MC events, produced as described in
Sect. 3.2, selected under the same criteria as the RD, and properly 
distributed over all the running periods.

Jets were reconstructed using the detected charged and neutral particles
of the event, the charged particles being selected by applying the following
criteria:
\begin{itemize}
\item  $p > 200$ MeV/$c$;
\item  $\Delta p/p < 100$\%;
\item  20$^{\circ} \leq \Theta \leq 160^{\circ}$;
\item track length $> 30$ cm;
\item impact parameters below 4 and 10 cm in the $R \Phi$ and $z$ projections,
respectively.
\end{itemize}
The neutral particles were taken within the geometrical acceptances of
the subdetectors in which they were reconstructed, within the selection
criteria of the appropriate subdetector pattern recognition codes
\cite{delphi1,delphi2}, without additional cuts. This effectively means 
that the detection threshold was about 400 MeV.

To reconstruct jets, the LUCLUS code \cite{luclus,luclus2,luclus3} with a fixed resolution
parameter $d_{join}=$ 3 GeV/$c$ was used. 
Only jets containing no 
identified electrons (positrons) and satisfying the condition
30$^{\circ} \leq \Theta_{jet} \leq$ 150$^{\circ}$ were taken for the analysis.
The minimum jet momentum was required to be 5 GeV/$c$. 

Photons were selected using the cuts described in Sect. 3.3.
A total of 694~530 converted photons was selected in the RD
and  2~368~641 converted photons in the MC.
%Of these, 164~890 RD photons and
%539~136 MC ones were in the selected $p_T$ region, $p_T < 80$ MeV/$c$.        

\section{Specifying the analysis variables}
\subsection{Signal definition}
As representatives of the photon rates the distributions of the $p_T$,
the photon transverse momentum with respect to the jet direction,
corrected for the detection efficiency were chosen (as mentioned above, 
only photons within the energy range of $0.2 - 1$ GeV are considered). 
To quantify the excess, the difference of the rates between the RD and MC
(the latter being normalized to the statistics of the RD events and corrected
by the recalibration procedure, see \cite{aspdel} and Sect.~6.2.2 of this 
work) was integrated in the $p_T$ interval from 0 to 80 MeV/$c$
and the value obtained was defined as a signal. 

\subsection{Jet momentum}
The jet momentum, $p_{jet}$, is defined as the vector sum of 3-momenta 
of all charged and neutral particles belonging to a given jet.
The distributions of this variable obtained with both, 
the real and the MC data, are shown in Fig. 1a. 
Due to uncertainties in the determination 
of the jet constituent momenta and lost particles, this variable is not 
accurately measured (which can be seen also from Fig. 1a, with the 
distributions showing the maximum at 40 GeV/$c$, shifted from the expected
value of 45.6 GeV/$c$, and a tail extending up to 60 GeV/c). In order
to evaluate the accuracy of the jet momentum reconstruction a comparison
of the $p_{jet}$ composed of the $measured$ particle momenta with the vector
sum of momenta of the $generated$ particles (i.e. before transporting them 
through the detector), assigned to a given jet, was done 
using the MC data. The assignment procedure was the following.
                                                                                
First, only stable and quasi-stable particles ($\pi^+$, $\pi^-$, $K^+$, 
$K^-$, $p$, $\bar p$, and muons) were selected among the generated 
charged particles, the selection cut (200 MeV/$c$) being applied
to them, similarly to the detected charged particles (see Sect. 4). 
Analogously, among the generated neutral particles only photons, 
$K^0_L$'s and neutrons were selected, 
imposing a cut at 400 MeV/$c$. Then for every selected 
$generated$ particle a loop on jets (found by LUCLUS with  $detected$ 
particles) was organized, calculating the generated particle opening angle 
to the jet axis. A generated particle has been assigned to that jet
to which its opening angle was minimal. Note, this assignment procedure was
also applied when defining, at the generator level, all the analysis 
variables described below.
                                                                                
The scatter plot of the reconstructed jet momenta, $p_{jet}$, versus jet 
momenta at the generator level determined via the procedure described above
is shown in Fig. 2a. It is seen that the plot is dominated
by the main diagonal (which corresponds to the equality of the generated and
measured jet momenta) up to about 30 GeV/$c$ where the accumulation
of events near the diagonal starts to 
spread (note, the measured jet momenta exceeding $p_{max} = 0.5 E_{cm}/c$ 
were reduced to that value in the plot, as well as in the analysis in general). 
The spread defined the bin size in the momentum variable employed 
in this work, chosen to be 5 GeV/$c$. 
To supply further information on the momentum bias,
the mean values of the reconstructed and generated momentum distributions 
in the individual $p_{jet}$ bins are given in Table 1 of the analysis section.
    
Closely related to the jet momentum, $p_{jet}$, is the jet energy $E_{jet}$,
which can be defined as the sum of the energies of jet particles (assuming
pion masses for them). This variable will not be used as an independent one,
entering however into the definitions of other variables, the jet mass and 
hardness, see Sect.~5.6.
%the jet Lorentz factor.

In what follows, all the jet variables (with one exception for the hardness 
$\kappa_J$) will be defined for jets having momenta
$p_{jet} > 20$~GeV/$c$ 
%thus taking into account the cut on the jet momenta introduced below 
(for the motivation of this cut see Sect.~7.1).

\subsection{Jet charged multiplicity}
The jet charged multiplicity, $N_{ch}$, is defined as the number of charged
particles measured in the DELPHI tracking system, as described in more detail 
in \cite{delphi2}, with the tracks satisfying the selection criteria listed 
in Sect. 4 and pertaining to a given jet. 
The distributions of this variable for the RD and MC are shown in Fig. 1b.
At the generator level, the jet
charged multiplicity, $N_{ch}^{gen}$, is defined as the number of stable
charged particles produced in the primary fragmentation or in the decays of 
particles with lifetimes shorter than $3\times 10^{-10}$ s which belong to 
a given jet. In particular, the charged particles from $K^0_s$ and $\Lambda$ 
decays were included in the $N_{ch}^{gen}$, irrespectively of how far from 
the interaction point the decay occurred, while the charged particles 
from $K^0_L$ decay were not.

As in the previous case, with the jet momenta, one faces the problem of 
associating an observed charged multiplicity of a jet $N_{ch}$ to the 
``true" one. Usually this problem is solved by making use of the multiplicity 
corrections with a matrix $P(j,i)$, whose elements, defined with the MC data, 
are the probabilities of a jet with observed charged multiplicity $j$ to have 
a ``true" charged multiplicity $i$ (the latter being determined via an 
assignment procedure analogous to that described in the previous section, 
this time for the generated charged particles only). Then the observed 
multiplicity is corrected accordingly to these probabilities. This method 
reproduces perfectly the distributions of the jet multiplicities, but being 
purely probabilistic it is not applicable when the jet multiplicity has to be 
used as an argument on a jet-by-jet basis. This requires a special 
consideration of the problem. 
%It was found with the MC events that the
%matrix corrections dilute the correlations between the rate of the inner
%hadronic bremsstrahlung and the jet charged multiplicity corrected in this 
%way, decreasing the dependence of the photon rate on the corrected $N_{ch}$
%variable as compared to the uncorrected one (a similar decrease was 
%observed also for the observed signal in the RD). Therefore in the current 
%study no corrections of the $N_{ch}$ were applied. 
 
%The comparison of the mean charged multiplicity of hadronic events at 
%$E_{cm} = m_Z$ quoted by the PDG to be $20.76 \pm 0.16$ \cite{multpdg}, 
%to the mean summary multiplicity of jets obtained in this study, 
%$<\sum_{j=1}^{N_jet} N_{ch}^j> = 18.6$, with a negligible statistical error,
%reveals the 10\% difference between the two numbers. 
%Thus  one can expect a systematic bias for 
%the $N_{ch}$ multiplicity variable, i.e. the bias of the $x$ axis values 
%in the soft photon rate dependences, to be, on average, about 10\%. 
%Though this bias was observed to be larger at low $N_{ch}$ multiplicities, 
%it was verified that it is always comparable to (generally, smaller than) the 
%$N_{ch}$ half bin widths employed, and therefore was considered to be 
%admissible in the current study. The validity of this assumption was tested 
%when the main results of a given analysis were obtained and used in a check
%procedure which is described in the Appendix.

The effect of the multiplicity migration is illustrated by Fig. 2b where
the scatter plot of the reconstructed jet $N_{ch}$ multiplicity versus
the jet charged multiplicity at the generator level is displayed. 
It shows the following features: the cells on the main 
diagonal are the most populated; non-diagonal elements are almost symmetric 
with respect to the diagonal, though some small prevalence of the 
under-diagonal terms relative to the above-diagonal ones can be seen, 
which corresponds to track losses. In order to keep the $N_{ch}$ systematic 
bias transparent, columns with the generated charged multiplicity mean values 
and their r.m.s. are given in Table 2 below to be compared with the $N_{ch}$. 
It can be seen that the bias of the $N_{ch}$ variable is always comparable to
(generally, smaller than) the $N_{ch}$ half bin widths employed, and therefore 
it was considered to be admissible in the current study. The validity of this 
assumption was tested when the main results of a given analysis were obtained 
and used to model the effect of the photon rate bias due to multiplicity
migration, with the effect being found to be negligible.                  
                                                               
\subsection{Jet neutral particle multiplicity}
The term ``neutral particle" specifies
a neutral hadron satisfying the selection criteria described below.
The criteria are aimed at counting neutral hadrons (which are $\pi^0$'s 
mainly) using the DELPHI electromagnetic calorimeter (HPC) and converted
photons. Also the DELPHI hadron calorimeter (HCAL) was used to detect a small
fraction of $K^0_L$'s and neutrons (antineutrons). The detected neutral showers 
(the showers which cannot be associated to any charged particle track) 
and converted photons were treated in the following way:
\begin{itemize}
 \item  HPC showers with the energy within the range from 1 to 6 GeV
were considered as photons; 
 \item  HPC showers with the energy exceeding 6 GeV were considered as
$\pi^0$'s;
 \item converted photons were collected if their energy exceeded 1 GeV;
 \item HCAL showers were collected if their energy exceeded 2 GeV, and
were considered as particles.
\end{itemize}

The jet neutral particle multiplicity, $N_{neu}$, has been defined then
as the number of its neutral particles, each photon being treated as a
half-particle, a $\pi^0$ constituent (in the case of half-integer $N_{neu}$
values, they were promoted to the next integers; it was tested that the effect 
of the $\pi^0$ overcounting due to ISR and $\pi^0$ photons entering different 
jets induced by this convention is negligible). For most of the events 
this effectively means that the neutral particle lower energy cut is 2 GeV.
The distributions of the $N_{neu}$, selected in such a way, are shown 
in Fig. 1c.

The same criteria were applied to the generated neutral particles. 
In particular, the photons from neutral pion decays were required to be 
within the HPC acceptance and to satisfy the energy cuts described above 
in order for the pion to be counted as a generated neutral particle.
                                                                                
The scatter plot of the reconstructed jet $N_{neu}$ multiplicity versus
the jet neutral multiplicity at the generator level is displayed in Fig. 2c. 
It shows features similar to those of the $N_{ch}$ plot: the prevalence
of the main diagonal elements and approximate symmetry of the non-diagonal
elements with respect to the diagonal. 

In order to keep the $N_{neu}$ systematic bias transparent,
columns with the generated neutral multiplicity mean values and their r.m.s.
are given in Table 3 below.
                                                                                
\subsection{Jet total particle multiplicity}
The term ``total particle multiplicity" (or simply ``particle
multiplicity"), $N_{par}$, denotes the sum of charged and neutral particles 
(as defined in the previous sections),
\begin{center}
          $N_{par} =$ $N_{ch} + N_{neu}$~,
\end{center}
and analogously for the multiplicities at the generator level.
The distributions of the  $N_{par}$ are shown in Fig. 1d.

The scatter plot of the reconstructed jet $N_{par}$ versus
the jet particle multiplicity at the generator level is shown in Fig. 2d.
Mean values of the generated particle multiplicities and their r.m.s.
are given in Table 4 below.

%\subsection{Jet mass and jet Lorentz factor}
\subsection{Jet mass and hardness}
The jet mass is defined as $M_{jet} c^2 = \sqrt{E_{jet}^2 - p_{jet}^2 c^2}$. 
The distributions of this variable are shown in Fig. 1e.
The scatter plot of the reconstructed jet mass $M_{jet}$ versus
the mass at the generator level is shown in Fig. 3a.

%Closely related to the jet mass is a variable $E_{jet}/M_{jet}$ which can
%be called the jet Lorentz factor, $\Gamma_{jet}$. 
%The scatter plot of the reconstructed jet Lorentz factor $\Gamma_{jet}$ 
%versus this variable at the generator level is shown in Fig. 3b.
 
The jet energy enters also in the variable which characterizes the hardness 
of the process producing the jet \cite{ddt1,ddt2}, $\kappa_J$, which is 
defined as follows:
\begin{center}
          $\kappa_J =$ $E_{jet} \sin \frac{\alpha}{2}$~,
\end{center}
where $\alpha$ is the angle to the closest jet. This variable corresponds 
to the beam energy in two-jet events without lost particles. In general, 
$\kappa_J$ depends on the topology of the event. Theoretical \cite{ddt1,ddt2} 
and experimental \cite{delhard1,delhard2} studies of hadron production in
processes with non-trivial topology have shown that characteristics of the
parton cascade depend essentially on this variable. Therefore it was involved
in the current analysis, restricting however this particular study to multi-jet
(three or more jets) events. A sample of 2~192~644 such events was selected
out of the total sample. The distributions of the $\kappa_J$ variable 
for these events are shown in Fig. 1f.

The scatter plot of the reconstructed $\kappa_J$ for jets with momenta 
$p_{jet} > 5$ GeV/$c$ versus this variable at the generator level is shown 
in Fig.~3b.

\subsection{Jet core characteristics}
\subsubsection{Jet core net charge}
The jet net charge, $Q_{net}$, is defined as the algebraic sum of the charges 
of the jet charged particles. Two kinds of jet net charges were tried,
as described in the next two paragraphs, respectively.

The first one was the ``raw" jet net charge, with all the jet particles
involved. No significant dependence of the soft photon production on this 
variable was found. Moreover, it has been known for a long time \cite{fife} 
that this variable is ill-defined since it fluctuates significantly depending 
on whether a positive (negative) soft particle is added to the jet or not. 
Therefore we leave it out of the presentation of the results. 

A more tractable quantity is the jet ``core" net charge, which was constructed
with those particles only, which had momenta exceeding 2 GeV/$c$ and were
confined within a cone of 100~mrad half-angle to the jet axis. 
The distributions of this variable are shown in Fig. 1g.
The scatter plot of the reconstructed jet core $Q_{net}$ versus 
this variable at the generator level is shown in Fig. 3c. 
It shows a good diagonal structure. 

In what follows, the absolute values of the core $Q_{net}$ 
will be used as the corresponding net charge. 
                    
\subsubsection{Jet core charged multiplicity}
The jet core charged multiplicity, core $n_{ch}$, was defined under the same
conditions as the previous variable, i.e. it is the number of jet charged
particles having momenta exceeding 2~GeV/$c$ and confined within a cone 
of 100~mrad half-angle to the jet axis.
The distributions of this variable are shown in Fig. 1h.
The scatter plot of the reconstructed jet 
core $n_{ch}$ versus this variable at the generator level is shown in Fig.~3d.
It shows a good diagonal structure.

\section{Treatment of systematic errors}
\subsection{General remarks}
In this section the treatment of systematic errors of the photon rates is 
described. First the systematic uncertainties in the determination of the
signal are defined and then those in the bremsstrahlung predictions.
The former can be subdivided into the uncertainties originating from the 
systematic effects biasing the MC distributions with respect to the RD ones 
(described in Sects. 6.2.1, 6.2.2 below) and the uncertainties common for both 
data sets (originating from the efficiency corrections, Sect. 6.2.3). 
%Such classification of the systematic uncertainties for the signal rate 
%is convenient, allowing the systematic effects induced by a possible imperfect 
%modelling of the DELPHI detector in the MC simulation (see \cite{aspdel}) 
%to be disentangled from those in the efficiency finding, to study them 
%separately, and to have as a result the efficiency tables (defined according 
%to the description given in Sect. 3.3) common for both the MC and RD.    
                                                                                
Before going into details of systematic error estimates an important remark
on the global values of systematic effects, which might stem from an excess 
of soft $\pi^0$'s and photons from their decays in the real data as compared to 
the MC, has to be made. These effects were tested in \cite{aspdel} by several 
methods and were found to be satisfactorily small. In particular, the test   
invoking almost precise $SU(2)$ symmetry of the strong interactions in order to 
use charged pions from hadronic decays of $Z^0$ for evaluation of the 
possible difference in production rates of neutral pions and, consequently, 
soft photons in the RD and MC, is described in Sect.~6.2 of \cite{aspdel}. 
With this test the expectations for the systematic 
bias of the photon rates in the RD and MC in the signal kinematic range were 
found to be below 10\% of the signal.   

Another test, described in Sect.~6.3 of \cite{aspdel}, involved 
the direct comparison of $\pi^0$ production in the RD and MC.
The upper limit for the systematic bias of the converted soft photon RD to 
MC ratio obtained from this test was below 20\% of the signal at 90\% CL.
Thus, the two tests agree and suggest that there is no substantial 
systematic effect due to the modelling and reconstruction of soft photons 
from $\pi^0$ decay.

The methods of the estimation of systematic uncertainties 
in the determination of the signal in individual bins of the variables 
under study are described below.  

\subsection{Systematic uncertainties in the determination of the signal}
\subsubsection{Event generator systematics}
This type of systematic effect arises mainly due to an improper reproduction 
of the experimental spectra of photons by the MC event generator, 
being a result of the modelling of the fragmentation process, 
i.e. parton shower and string hadronization as implemented in 
\cite{jetset,jetset2,jetset3}. Another generator systematic bias which can be
induced by an inadequate representation of the full set of unstable hadrons
decaying radiatively (other than $\pi^0$'s) at the final stage of the 
hadronization mechanism was carefully studied in 
\cite{aspdel}, and its uncertainty was shown to be small as compared to 
other components of the systematics; thus it will be neglected in this study. 
                                                                                
The systematic errors due to the JETSET fragmentation model and its tuning 
were estimated in two steps. First, the MC data produced with three 
different tunings described in \cite{tun1,tun1a,tun2} were analyzed separately 
in order to extract the systematic error due to the generator tuning. 
Comparing the photon spectra in the individual bins of variables under study 
listed in the previous section, this component of the systematic error was 
determined for every bin of the variables as the r.m.s. of the soft photon 
rates in the $p_T < 80$ MeV/$c$ region.

Then the MC data produced with ARIADNE were studied.
Comparing the photon spectra produced with this generator to those of  
JETSET, the systematic uncertainties due to the generator model for the 
rate of soft photons of $p_T < 80$ MeV/$c$ were evaluated for each bin 
of the variables under study as half the difference between the JETSET 
and ARIADNE rates.

The typical individual bin systematic error due to the generator 
was found to be at the level of about (15 - 25)\% of the signal, 
the main contribution to this error coming from the generator model component. 

\subsubsection{Detector systematics}
This type of systematic effect (called hardware systematics in 
\cite{aspdel}) is related to biases in the simulation of the detector and
experimental conditions in the MC stream, i.e. those which appear
when transporting MC photons through the DELPHI setup and reconstructing 
them (after conversion in the DELPHI setup material) from hits simulated 
in the TPC. These features have been extensively studied in \cite{aspdel},
and a recalibration procedure was elaborated in order to reduce this bias.
It used wide angle photons, $\theta_{\gamma} > $ 200 mrad (keeping 
the $E_\gamma < 1$ GeV), for which the signal of the direct soft photons was  
assumed to be zero, to re-normalize the material distribution along the
photon path in the simulation, and to account for possible differences in 
reconstruction of converted photons from the TPC hits along $e^+ e^-$ tracks 
in the MC and RD. 

The recalibration was applied to each individual bin of the 
variables described in Sect.~5. Varying the recalibration parameters and the
MC data samples, the resulting detector systematic errors for the signal were 
found to be at the level of about 10\%  of the signal (on average) after the 
recalibration. 

\subsubsection{Systematic errors due to efficiency corrections}
The systematic errors due to the method of implementation of the efficiency 
correction in the photon $p_T$ range below 80 MeV/$c$ were determined 
individually for every bin of the variables under current study from the MC data
and consisted of two components. The first one, induced by an 
interpolation method, is described in Sect. 3.3. The second component of this 
error is a purely instrumental effect originating from the 
conversion method resolution in energy and the efficiency binning over this 
variable. It was estimated by comparing the photon $p_T$ distributions 
taken at the output of the event generator to the analogous distributions 
of the photons (after they had been transported through the DELPHI detector 
by DELSIM with a subsequent simulation of their conversions) corrected for 
efficiency. 
Both components of the systematic error under discussion were combined 
in quadrature, resulting in a typical value of the individual bin errors 
induced by efficiency corrections to be at the level of about 6\% of the signal.

These errors were summed quadratically with other components of the signal
systematic error described above, thus giving the overall systematic 
uncertainty in the finding of the signal. However, 
due to strong bin-to-bin correlations of the systematic errors
we will not use them in what follows when fitting signal dependence curves.
  
\subsection{Systematic uncertainty of the bremsstrahlung predictions}
The systematic uncertainty for the bremsstrahlung predictions resulting from
formula (2) was estimated by comparing the ISR rates obtained with this formula
and those delivered by the DYMU3 generator in the photon $p_T$ 
range (defined to the beam direction) below 80 MeV/$c$, as the difference 
between the predictions. This difference was about 4\%, and it was taken as
the systematic error for the bremsstrahlung predictions. 

This value was found to be close to the difference in predictions 
for the inner hadronic bremsstrahlung rate obtained with formula (2), 
and those calculated taking into account the higher order radiative 
corrections, by the use of an exponentiated photon spectrum \cite{hais}. 
When doing these calculations, the $\beta$ which governs 
the bremsstrahlung photon spectrum was obtained by integration of formula (3) 
in \cite{hais} applying the $p_T$ cut imposed by the signal definition, $p_T<$ 
0.08 GeV/$c$, i.e. within a rather narrow angular range varying as a function 
of the photon energy according to the aforementioned cut. The $\beta$ values 
were found to be 0.0106 and 0.0135 with the minimum jet momentum cut 
at 5 and 20 GeV/$c$, respectively, which would lead in both cases to less than 6\% 
difference between formula (2) and exponentiation method predictions.

Another component of the bremsstrahlung prediction error originating from the 
uncertainties in the charged particle spectra coming from the event 
generator was determined in \cite{aspdel} to be 5\%  
by varying JETSET tunings and comparison with ARIADNE.

\section{Results of the analysis}
Throughout this section the dependences of the direct soft photon production 
on the jet variables will be considered in comparison with those of the inner 
hadronic bremsstrahlung (often referring to these dependences as to the signal 
and bremsstrahlung behaviour). 
The overall excess factor of about four over the bremsstrahlung, 
which can be easily seen in all the tables and plots below,  will be taken 
for granted in the following, even when not explicitly mentioned.
   
\subsection{Signal dependence on jet momentum}
The distribution of the signal rate against the jet momentum is tabulated 
in Table 1 and plotted in Fig. 4 (left panel), together with the corresponding
predictions for the inner hadronic bremsstrahlung rates. In this figure 
(as well as in the following ones) the inner vertical bars represent the
statistical errors, while the whole vertical bars give the statistical and
systematic errors combined in quadrature. The inner hadronic bremsstrahlung 
predictions are shown in the figure by triangles \footnote{Note, 
the bremsstrahlung rates given throughout this Section are calculated within 
the bins defined with the detected (not generated) variables.}.  

As can be seen from the 
figure, the signal rate first increases with the jet momentum, similarly to 
the predicted bremsstrahlung rate, then it stops increasing, within the
errors, at jet momenta about 20 GeV/$c$. The bremsstrahlung 
rate also shows a tendency to a saturation with the momentum increase. The
curve fitting the bremsstrahlung points is a polynomial of the 2nd order. 
The same curve scaled by a factor of 3.9 is drawn through the signal points.
It describes them well. The ratio of the signal rate to that 
of the predicted bremsstrahlung is displayed in the right panel of Fig. 4
showing a uniform distribution. 
 
\vskip 0.4cm
{\bf Table 1.} The dependence of direct soft photon rates on the jet momentum. 
The first errors are statistical, the second ones are systematic.
\vskip 0.4cm
\begin{center}
\begin{tabular}{ |c c c c c| }
\hline
&&&&\\
$~~p_{jet},~~$&$~~~<p_{jet}>,~~~$&$<p_{jet}^{gen}>,$&~~~~~~~~~~~~~Signal,~~~~~~~~~~~~~&~~~~~Bremsstrahlung,~~~~~~~\\
GeV/$c$      &GeV/$c$            &GeV/$c$           &      $10^{-3} \gamma$/jet       &    $10^{-3} \gamma$/jet   \\
&&&&\\
\hline
&&&&\\
~~5$-$10&~~7.5 &$ ~8.6 $&$ ~~25\pm ~7\pm~~9$&$~~5.8\pm 0.1\pm 0.4$\\
 10$-$15& 12.4 &$ 13.6 $&$ ~~35\pm 10\pm 10$&$ 11.2\pm 0.1\pm 0.7$\\
 15$-$20& 17.5 &$ 19.6 $&$ ~~68\pm 12\pm 17$&$ 15.6\pm 0.1\pm 1.0$\\
 20$-$25& 22.5 &$ 25.7 $&$ ~~95\pm 11\pm 15$&$ 18.4\pm 0.1\pm 1.2$\\
 25$-$30& 27.6 &$ 30.9 $&$ ~~93\pm 10\pm 17$&$ 20.2\pm 0.1\pm 1.3$\\
 30$-$35& 32.5 &$ 34.8 $&$ ~~83\pm ~9\pm 16$&$ 22.2\pm 0.1\pm 1.4$\\
 35$-$40& 37.5 &$ 37.6 $&$  102\pm ~9\pm 17$&$ 24.4\pm 0.1\pm 1.6$\\
 40$-$45& 43.4 &$ 40.1 $&$ ~~75\pm ~6\pm 19$&$ 23.8\pm 0.1\pm 1.5$\\
&&&&\\
\hline
\end{tabular}
\end{center}
\vskip 0.4cm

In what follows, only jets with momenta exceeding 20 GeV/$c$ will be taken
for the analysis (with one exception for the hardness variable, $\kappa_J$). 
Though general tendencies of the signal behaviour with and without this
cut are similar, the cut is made in order to separate the momentum and 
other variable dependences, making use of the weakness of the photon 
production rate dependence on the jet momenta at $p_{jet} \geq$ 20 GeV/$c$ as 
noticed in the previous paragraph. The integral production rate of direct soft 
photons obtained with this cut is 
$(86.3 \pm 4.1 \pm 19.5) \times 10^{-3} \gamma$/jet, while the calculated 
bremsstrahlung rate is $(21.70 \pm 0.02 \pm 1.39) \times 10^{-3} \gamma$/jet.
This can be compared to the photon rates obtained without momentum cut, 
$(69.1 \pm 4.5 \pm 15.7) \times 10^{-3} \gamma$/jet for the signal and
$(17.10 \pm 0.01 \pm 1.21) \times 10^{-3} \gamma$/jet for the inner
hadronic bremsstrahlung, which were reported in \cite{aspdel}.  

\subsection{Signal dependence on jet charged multiplicity} 
The signal dependence on the jet charged multiplicity, as defined 
in Sect. 5.3, is tabulated in Table 2 and displayed in Fig. 5. 
As found with the MC $\tau^+ \tau^-$ events, the $\tau$ channel 
contaminations (as a fraction of the signal rates in the corresponding 
bins) in this distribution are:
\begin{itemize}
 \item $(0.3 \pm 0.1)\%$ in the $N_{ch} = 0-2$ bin; 
 \item $(0.1 \pm 0.1)\%$ in the $N_{ch} = 3,4$ bin; 
 \item $(0.3 \pm 0.1)\%$ in the $N_{ch} = 5$ bin; 
 \item $< 0.2\%$ (at 95\% CL) in subsequent bins. 
\end{itemize}

As can be seen from Fig. 5, the observed signal rate dependence follows 
in general, by a scale factor of about 4, that of the hadronic 
bremsstrahlung, though there is some excess in the first $N_{ch}$ bin.
The curve fitting the bremsstrahlung points in the left panel of Fig. 5 
is a 2nd order polynomial.
% with the constant term found to be $(8.6 \pm 0.1)\times 10^{-3} \gamma$/jet.  
%It can be compared to the predicted bremsstrahlung rate 
%for the $measured$ $N_{ch}=0$ (at which however the $generated$ charged 
%multiplicity is not zero on average, being $<N_{ch}^{gen}>=1.38 \pm 1.65$)
%which is $8.4 \times 10^{-3} \gamma$/jet. 
%In this way the charged particle losses leading to a situation when the 
%predicted inner hadronic bremsstrahlung rate is not vanishing while no 
%charged particles are detected in the jet, are reflected in the fit.
%are accounted for in the fit. 
The same curve scaled by a factor of 4 is drawn through the signal points.
It describes them satisfactorily, except perhaps the first point.
A similar conclusion can be drawn from the plot in the right panel of Fig. 5 
in which the ratio of the signal and predicted bremsstrahlung rates 
is displayed.
%\footnote{A general trend of the signal rate
%to grow with the $N_{ch}$, which can be seen in Fig. 5, does not 
%contradict the constant ratios of the RD to MC photon rates when varying 
%the charged multiplicity, $N_{ch}$, which were reported in \cite{aspdel} 
%(Sect. 6.4 and Table 5 in \cite{aspdel}). These observations, taken together, 
%just mean that the overall soft photons rates (signal $+$ background) in 
%the high $N_{ch}$ multiplicity jets are higher as compared to those in 
%the low $N_{ch}$ multiplicity jets to approximately the same extent in 
%the RD and MC.}. 

Note the muon bremsstrahlung point (an asterisk at the position $N_{ch} = 1$) 
in the right panel of Fig. 5. It is placed there using the results of 
the paper \cite{muonbrems} in which a good agreement of the observed inner 
bremsstrahlung from muons of dimuon events of $Z^0$ decays with 
the QED predictions is reported: the ratio of the observed direct 
soft photon production rate to the predicted level of the muon inner 
bremsstrahlung was found in \cite{muonbrems} to be $1.06 \pm 0.13 \pm 0.06$. 
This defined the ordinate of the muon point on the plot.

\vskip 0.4cm
{\bf Table 2.} The dependence of direct soft photon rates on the jet charged
multiplicity. 
\begin{center}
\begin{tabular}{ |c c c c c c| }
\hline
&&&&&\\
$~~N_{ch}~~$&$~<N_{ch}>$&$<N_{ch}^{gen}>$&$<N_{ch}^{gen}>$&~~~~~~~~Signal,~~~~~~~~~~&~~~~Bremsstrahlung,~~~~~\\
            &                &                &     r.m.s.     &  $10^{-3} \gamma$/jet      &   $10^{-3} \gamma$/jet   \\
&&&&&\\
\hline
&&&&&\\
0$-$2  &~1.68& ~2.15 &  1.07 &$ 97\pm  19\pm 20$&$ 12.1\pm 0.1\pm 0.8$\\
 3,4   &~3.63& ~4.01 &  1.15 &$ 65\pm  ~9\pm 12$&$ 17.0\pm 0.1\pm 1.1$\\
  5    &~5.00& ~5.30 &  1.17 &$ 67\pm  10\pm 14$&$ 19.3\pm 0.1\pm 1.2$\\
  6    &~6.00& ~6.22 &  1.25 &$ 83\pm  10\pm 18$&$ 20.9\pm 0.1\pm 1.3$\\
  7    &~7.00& ~7.15 &  1.33 &$ 90\pm  11\pm 18$&$ 22.7\pm 0.1\pm 1.4$\\
 8,9   &~8.45& ~8.45 &  1.52 &$ 93\pm  ~9\pm 20$&$ 24.8\pm 0.1\pm 1.6$\\
10,11  &10.41& 10.22 &  1.67 &$110\pm  13\pm 21$&$ 27.3\pm 0.1\pm 1.7$\\
12$-$16&13.19& 12.68 &  2.27 &$139\pm  17\pm 24$&$ 29.2\pm 0.1\pm 1.9$\\
&&&&&\\
\hline
\end{tabular}
\end{center}
\vskip 0.2cm
 
\subsection{Signal dependence on jet neutral particle multiplicity} 
The signal dependence on the jet neutral multiplicity, as defined in 
Sect. 5.4, is tabulated in Table 3 and shown in Fig. 6.

The contamination from $Z^0 \rightarrow \tau^+ \tau^-$ events in the various
$N_{neu}$ bins was found to be (as a fraction of the signal rates in the
corresponding bins): 
\begin{itemize}
 \item $(1.3 \pm 0.3)\%$ in the $N_{neu} = 0  $ bin;
 \item $(0.4 \pm 0.1)\%$ in the $N_{neu} = 1  $ bin;
 \item $(0.2 \pm 0.1)\%$ in the $N_{neu} = 2$ bin; 
 \item $< 0.2\%$ (at 95\% CL) in subsequent bins.
\end{itemize}

\vskip 0.4cm
{\bf Table 3.} The dependence of direct soft photon rates on the jet neutral 
multiplicity. 
\begin{center}
\begin{tabular}{ |c c c c c c| }
\hline
&&&&&\\
$~~N_{neu}~~$&$~<N_{neu}>$&$<N_{neu}^{gen}>$&$<N_{neu}^{gen}>$&~~~~~~~~~Signal,~~~~~~~~~~~~&~~Bremsstrahlung,~~~~~\\
             &                 &            &   r.m.s.        &      $10^{-3} \gamma$/jet    &    $10^{-3} \gamma$/jet\\
&&&&&\\
\hline
&&&&&\\
 0   & 0  &  0.53  &  0.74 &$ ~41\pm 11\pm 16$&$ 22.3\pm 0.1\pm 1.4$\\
 1   & 1  &  1.20  &  0.91 &$ ~59\pm ~7\pm 14$&$ 22.7\pm 0.1\pm 1.4$\\
 2   & 2  &  2.15  &  1.37 &$ ~99\pm ~7\pm 17$&$ 21.1\pm 0.1\pm 1.3$\\
 3   & 3  &  2.92  &  1.50 &$ 115\pm 10\pm 24$&$ 19.1\pm 0.1\pm 1.2$\\
 4   & 4  &  3.66  &  1.70 &$ 175\pm 18\pm 31$&$ 17.1\pm 0.1\pm 1.1$\\
5$-$7&5.18&  4.25  &  1.77 &$ 226\pm 38\pm 48$&$ 13.6\pm 0.1\pm 0.9$\\
&&&&&\\
\hline
\end{tabular}
\end{center}
\vskip 0.4cm

As can be seen from Fig. 6, the signal behaviour differs drastically 
from that of the inner hadronic bremsstrahlung predictions. 
A possible interpretation of this difference will be given in Sect. 8, when
considering various theoretical approaches to the problem of the soft photon
excess in reactions of multiple hadron production.

\vskip 0.3cm
\subsection{Signal dependence on the jet total particle multiplicity}
The signal dependence on the jet particle multiplicity is given in
Table 4 and shown in Fig. 7. The contamination from 
$Z^0 \rightarrow \tau^+ \tau^-$ events in the various particle 
multiplicity bins was found to be (as a fraction of the signal 
rates in the corresponding bins):
\begin{itemize}
 \item $(0.7 \pm 0.2)\%$ in the $N_{par} = 1-4$ bin; 
 \item $(0.3 \pm 0.1)\%$ in the $N_{par} = 5,6$ bin; 
 \item $(0.2 \pm 0.1)\%$ in the $N_{par} = 7$ bin; 
 \item $(0.6 \pm 0.1)\%$ in the $N_{par} = 8$ bin; 
 \item $(0.9 \pm 0.1)\%$ in the $N_{par} = 9$ bin; 
 \item $< 0.1\%$ (at 95\% CL) in subsequent bins. 
\end{itemize}

As can be seen from Fig. 7, the signal behaviour differs from that of the predicted bremsstrahlung, similarly to the previous case.
The discussion of this difference is given in Sect. 8. Here we note only that the signal distribution can be fitted satisfactorily
by a straight line passing through the origin of the coordinate system,
as shown in the left panel of Fig.~7.

\vskip 4mm
{\bf Table 4.} The dependence of direct soft photon rates on the jet total 
particle multiplicity. 
\begin{center}
\begin{tabular}{ |c c c c c c| }
\hline
&&&&&\\
$~~N_{par}~~$&$~<N_{par}>$&$<N_{par}^{gen}>$&$<N_{par}^{gen}>$&~~~~~~~~~~Signal,~~~~~~~~~~~&~~Bremsstrahlung,~~~~~~\\
             &                 &            &     r.m.s.      &     $10^{-3} \gamma$/jet     &   $10^{-3} \gamma$/jet \\
&&&&&\\
\hline
&&&&&\\
 1$-$4 & ~3.50& ~4.24 &  1.66 &$ ~45\pm 16\pm 12$&$ 18.3\pm 0.1\pm 1.2$\\
  5,6  & ~5.60& ~6.19 &  1.72 &$ ~51\pm ~9\pm 11$&$ 19.2\pm 0.1\pm 1.2$\\
   7   & ~7.00& ~7.46 &  1.74 &$ ~77\pm 10\pm 17$&$ 20.3\pm 0.1\pm 1.3$\\
   8   & ~8.00& ~8.35 &  1.80 &$ ~77\pm 10\pm 21$&$ 21.6\pm 0.1\pm 1.4$\\
   9   & ~9.00& ~9.23 &  1.85 &$ ~81\pm 11\pm 20$&$ 23.2\pm 0.1\pm 1.5$\\
 10,11 & 10.45& 10.50 &  1.96 &$ 110\pm ~9\pm 24$&$ 24.7\pm 0.1\pm 1.6$\\
 12,13 & 12.41& 12.20 &  2.06 &$ 138\pm 14\pm 26$&$ 27.0\pm 0.1\pm 1.7$\\
14$-$17& 15.16& 14.59 &  2.49 &$ 167\pm 18\pm 30$&$ 28.7\pm 0.1\pm 1.8$\\
&&&&&\\
\hline
\end{tabular}
\end{center}
\vskip 0.3cm

\subsection{Signal rates in the 2-dimensional distribution 
{\boldmath $N_{ch}$ vs $N_{neu}$}}
Due to $SU(2)$ symmetry of the strong interactions and/or selection cuts, 
the variables $N_{ch}$ and $N_{neu}$ can be correlated. 
%\footnote{For an example of such an (anti)correlation see Sect. 7.7.}. 
In order to disentangle the signal 
rate dependences on these variables, the two-dimensional signal distribution 
as a function of the $N_{ch}$ and $N_{neu}$ was studied. When doing this,
the range of the jet polar angles $\Theta_{jet}$ to the beam was 
restricted to the interval of 50$^{\circ} \leq\Theta_{jet}\leq 130^{\circ}$.
This restriction equalizes, practically, the angular acceptances for the 
charged and neutral particles, the latter being mainly $\pi^0$'s detected
by the HPC via their decay photons. This equalization is important when 
comparing quantitatively the photon rate dependences on the above variables.
For the same reason (to equalize detection efficiencies for charged and 
neutral particles) a lower momentum cut at 2 GeV/$c$ was introduced when
calculating the charged particle multiplicity for this particular analysis.

The signal rates obtained with this selection are given as a two-dimensional 
distribution presented in Table 5 \footnote{The rates in the 1st and 5th lines 
of the signal column in Table 5 were corrected for the effect induced by the
cut $p_{jet} \geq$ 20 GeV/$c$ after appropriate study of the influence of this 
cut on the signal rates at small $N_{ch}$ multiplicities, see comment on
this influence given in Sect. 7.7.}. 

The distribution was fitted by the simplest possible form
$ R = a_1 N_{ch} + a_2 N_{neu}$ with a reasonable value of the
reduced $\chi^2$ close to 1 (the statistical errors only being
used in the fit). The values of the fitted rates are given in the last column
of Table 5. The linear dependence coefficients $a_1$ and $a_2$ obtained with
the fit are $(6.9 \pm 1.8 \pm 1.8) \times 10^{-3} \gamma$/jet
and $(37.7 \pm 3.0 \pm 3.6) \times 10^{-3} \gamma$/jet, respectively.
The first errors of these values are the fit parameter errors based on the
statistical errors of the signal rates. The second errors represent the fit
parameter changes obtained by adding to the signal rate central values
their systematic errors taken randomly accordingly to a Gaussian distribution,
and repeating this procedure many times to find at the end the r.m.s. of the
fit parameters; in this way a propagation of the systematic uncertainties
of the rates to those of the fit parameters was performed.

A straightforward conclusion which can be drawn from the difference of the
two coefficients is that the dependence of the signal rates on the
jet neutral multiplicity is a factor of $5.4 \pm 1.4 \pm 1.5$
stronger than the dependence on the charged multiplicity \footnote
{We have tested that the systematic error of the quoted factor is valid
also in the case of strong bin-to-bin correlations of systematic errors
of individual photon rates.}.
A possible interpretation of this difference is suggested in Sect. 8.

\vskip 0.2cm
{\bf Table 5.} Two-dimensional dependence of direct soft photon rates on the 
jet multiplicities, $N_{ch}$ versus $N_{neu}$. 
\begin{center}
\begin{tabular}{ |c c c c c c| }
\hline
&&&&&\\
$~~N_{ch}~~$&$~<N_{ch}>$&$<N_{ch}^{gen}>$&$<N_{ch}^{gen}>$&~~~~~~~~~~~~~~Signal,~~~~~~~~~~~~~~&~~~~~Fit,~~~~~~~~~\\
              &                &           &   r.m.s.       &      $10^{-3} \gamma$/jet       &$10^{-3} \gamma$/jet~~\\
&&&&&\\
\hline
&&&&&\\
\multicolumn{6}{|c|}{jet $N_{neu}=0,1, <N_{neu}>=0.76$ }\\
&&&&&\\
 0,1 & 0.91 &  1.27 &  0.82 &$ 47 \pm 14\pm ~9$&$ 37 $\\
 2,3 & 2.61 &  2.87 &  0.93 &$ 48 \pm ~9\pm 11$&$ 49 $\\
 4,5 & 4.38 &  4.49 &  0.99 &$ 63 \pm 10\pm 12$&$ 61 $\\
6$-$9& 6.42 &  6.26 &  1.14 &$ 79 \pm 21\pm ~9$&$ 76 $\\
&&&&&\\
\multicolumn{6}{|c|}{jet $N_{neu}=2$ }\\
&&&&&\\
 0,1 & 0.85 &  1.27 &  0.84 &$ ~99\pm 30\pm 17$&$ ~81 $\\
 2,3 & 2.55 &  2.86 &  0.93 &$ 116\pm 29\pm 31$&$ ~93 $\\
 4,5 & 4.36 &  4.52 &  0.99 &$ ~93\pm 18\pm 27$&$ 106 $\\
6$-$9& 6.38 &  6.33 &  1.08 &$ 121\pm 39\pm 19$&$ 120 $\\
&&&&&\\
\multicolumn{6}{|c|}{jet $N_{neu}=3$ }\\
&&&&&\\
 0,1 & 0.80 &  1.24 &  0.87 &$ 111\pm 26\pm 19$&$ 119 $\\
 2,3 & 2.51 &  2.83 &  0.93 &$ ~97\pm 17\pm 29$&$ 130 $\\
 4,5 & 4.34 &  4.52 &  0.96 &$ 155\pm 25\pm 48$&$ 143 $\\
6$-$9& 6.31 &  6.29 &  1.05 &$ 234\pm 61\pm 32$&$ 157 $\\
&&&&&\\
\multicolumn{6}{|c|}{jet $N_{neu}=4-6, <N_{neu}>=4.40$ }\\
&&&&&\\
 0,1 & 0.75 &  1.20 &  0.89 &$ 207\pm 24\pm 35$&$ 171 $\\
 2,3 & 2.46 &  2.81 &  0.95 &$ 170\pm 21\pm 47$&$ 183 $\\
 4,5 & 4.29 &  4.50 &  0.96 &$ 205\pm 31\pm 62$&$ 196 $\\
6$-$9& 6.25 &  6.24 &  1.06 &$ 256\pm 91\pm 94$&$ 209 $\\  
&&&&&\\
\hline
\end{tabular}
\end{center}
\vskip 0.2cm

%\subsection{Signal dependence on jet mass and jet Lorentz factor}
\subsection{Signal dependence on jet mass and hardness}
%The dependences of the signal rate on the jet mass and the jet Lorentz factor
%are given in Tables 6,7 and shown in Figs. 8,9, respectively. As can be seen 
%from these figures, both the signal and the predicted bremsstrahlung rates 
%behave similarly, depending rather weakly on these jet characteristics.
%Thus the jet mass and the jet Lorentz factor turn out to be not a very 
%suitable variables in order to be used for the direct soft photon production 
%rate parameterization. Nevertheless, 
%the observed behaviour of the production rates against these variables 
%(the rate flatness) can supply useful information when comparing the 
%various soft photon production models discussed in Sect. 8.
 
The dependence of the signal rate on the jet mass 
is given in Table~6 and shown in Fig.~8. As can be seen
from the figure, both the signal and the predicted bremsstrahlung rates
behave similarly, depending rather weakly on this jet characteristic.
Thus the jet mass turns out to be not a very suitable
variable to use for the direct soft photon production rate
parameterization. Nevertheless,
the observed behaviour of the production rates against this variable
(the rate flatness) can supply useful information when comparing the
various soft photon production models discussed in Sect.~8.

\vskip 0.3cm
{\bf Table 6.} The dependence of direct soft photon rates on the jet mass. 
\begin{center}
\begin{tabular}{ |c c c c c| }
\hline
&&&&\\
$M_{jet},$&$~~<M_{jet}>,~~$&$<M_{jet}^{gen}>,$&~~~~~~~~~~~Signal,~~~~~~~~~~~&~~~~~Bremsstrahlung,~~~~~~\\
 GeV/$c^2$&   GeV/$c^2$    &   GeV/$c^2$      &      $10^{-3} \gamma$/jet   &     $10^{-3} \gamma$/jet \\
&&&&\\
\hline
&&&&\\
 ~1.5$-$~3.0 & ~2.44 & $~3.51$ &$ ~79\pm 15\pm 12$ & $ 20.4\pm 0.2\pm 1.3$ \\
 ~3.0$-$~4.5 & ~3.83 & $~4.59$ &$ ~82\pm ~9\pm 16$ & $ 20.9\pm 0.2\pm 1.3$ \\
 ~4.5$-$~6.0 & ~5.27 & $~5.82$ &$ ~73\pm ~7\pm 19$ & $ 21.3\pm 0.2\pm 1.4$ \\
 ~6.0$-$~7.5 & ~6.73 & $~7.07$ &$ ~84\pm ~7\pm 17$ & $ 21.7\pm 0.2\pm 1.4$ \\
 ~7.5$-$~9.0 & ~8.20 & $~8.30$ &$ ~89\pm ~8\pm 18$ & $ 22.4\pm 0.2\pm 1.4$ \\
 ~9.0$-$10.5 & ~9.67 & $~9.45$ &$ 109\pm 10\pm 17$ & $ 22.6\pm 0.2\pm 1.4$ \\
 10.5$-$12.0 & 11.14 & $10.49$ &$ ~91\pm 14\pm 16$ & $ 22.1\pm 0.2\pm 1.4$ \\
 12.0$-$15.0 & 13.10 & $11.00$ &$ 134\pm 20\pm 18$ & $ 22.2\pm 0.2\pm 1.4$ \\
&&&&\\
\hline
\end{tabular}
\end{center}
\vskip 0.3cm

%\vskip 0.2cm
%{\bf Table 7.} The dependence of direct soft photon rates on the jet Lorentz 
%factor. 
%\begin{center}
%\begin{tabular}{ |c c c c c| }
%\hline
%&&&&\\
%$~~~\Gamma_{jet}~~~$&$~~~~<\Gamma_{jet}>~~~$&$<\Gamma_{jet}^{gen}>$&~~~~~~~~~~~~Signal,~~~~~~~~~~&~~~~~Bremsstrahlung,~~~\\
%                    &                       &                      &            $10^{-3} \gamma$/jet&  $10^{-3} \gamma$/jet    \\
%&&&&\\
%\hline
%&&&&\\
% 2$-$4 & 3.48 &$ 3.85$&$  98\pm  9\pm 17$&$ 19.4\pm 0.1\pm 1.2$\\
% 4$-$5 & 4.50 &$ 4.68$&$  93\pm  7\pm 19$&$ 21.7\pm 0.1\pm 1.4$\\
% 5$-$6 & 5.47 &$ 5.51$&$  79\pm  8\pm 17$&$ 22.1\pm 0.1\pm 1.4$\\
% 6$-$7 & 6.46 &$ 6.31$&$  82\pm 10\pm 16$&$ 22.2\pm 0.1\pm 1.4$\\
% 7$-$8 & 7.46 &$ 7.08$&$  85\pm 12\pm 18$&$ 22.3\pm 0.2\pm 1.4$\\
% 8$-$9 & 8.46 &$ 7.86$&$  61\pm 15\pm 17$&$ 22.4\pm 0.1\pm 1.4$\\
% 9$-$11& 9.86 &$ 8.85$&$  88\pm 15\pm 15$&$ 21.9\pm 0.1\pm 1.4$\\
%11$-$16& 13.02&$11.20$&$  83\pm 16\pm 12$&$ 21.7\pm 0.1\pm 1.4$\\
%&&&&\\
%\hline
%\end{tabular}
%\end{center}

Let us turn now to the hardness variable treated in multi-jet events.
As mentioned in Sect. 5.2, the jet momentum cut at 20 GeV/$c$ was not
applied when selecting jets for this particular analysis since the motivation 
for this cut is not justified in the given case. 
The integral production rate of direct soft photons obtained with these events
is $(63.7 \pm 4.0 \pm 13.9) \times 10^{-3} \gamma$/jet, 
while the calculated hadron bremsstrahlung rate is predicted to be 
$(15.80 \pm 0.01 \pm 1.01) \times 10^{-3} \gamma$/jet. The observed 
photon rates (see Table~7 and Fig.~9) show a fast increase with $\kappa_J$ 
in the first bins of this variable, followed by a saturation effect 
above 20 GeV, and they can be fitted satisfactorily by the 
bremsstrahlung curve scaled by a factor of~4. This means that the 
observability of a dependence of the direct soft photon production on the
hardness, which might differ from the analogous dependence of the hadron 
bremsstrahlung, is below the sensitivity of our approach.  

Very similar results were obtained with 2-jet events included in the analysis.
 
\vskip 0.4cm
{\bf Table 7.} The dependence of direct soft photon rates on the hardness of
the process producing the jet, $\kappa_J$. 
\begin{center}
\begin{tabular}{ |c c c c c| }
\hline
&&&&\\
~~~~$\kappa_J$~~~~~&~~~~~~~$<\kappa_J>$~~~~~~&$<\kappa_J^{gen}>$&~~~~~~~~~~~~Signal,~~~~~~~~~~~~&~~~~~Bremsstrahlung,~~~~\\
   GeV      &    GeV         &    GeV       &     $10^{-3} \gamma$/jet       &  $10^{-3} \gamma$/jet    \\
&&&&\\
\hline
&&&&\\
 ~2$-$5  & ~3.5 & ~3.7 & $ ~38\pm ~7\pm 15$ & $ 11.0\pm 0.1\pm 0.7$ \\
 ~5$-$10 & ~7.1 & ~7.0 & $ ~63\pm ~8\pm 18$ & $ 14.6\pm 0.1\pm 0.9$ \\
 10$-$15 & 12.2 & 12.2 & $ ~87\pm 12\pm 15$ & $ 17.3\pm 0.1\pm 1.1$ \\
 15$-$20 & 17.4 & 17.6 & $ 108\pm 16\pm 21$ & $ 18.5\pm 0.1\pm 1.2$ \\
 20$-$25 & 22.5 & 23.3 & $ ~68\pm 16\pm 17$ & $ 19.7\pm 0.1\pm 1.3$ \\
 25$-$30 & 27.5 & 28.2 & $ ~90\pm 16\pm 16$ & $ 21.2\pm 0.1\pm 1.4$ \\
 30$-$35 & 32.5 & 32.2 & $ ~74\pm 16\pm 15$ & $ 23.1\pm 0.1\pm 1.5$ \\
 35$-$40 & 37.5 & 36.0 & $ ~81\pm 16\pm 13$ & $ 24.8\pm 0.1\pm 1.6$ \\
&&&&\\
\hline
\end{tabular}
\end{center}
\noindent

\vskip 0.4cm
\subsection{Signal dependence on the jet core characteristics}
The dependences of the signal rate on the jet core net charge and the jet 
core charged multiplicity are given in Tables 8,9 and shown in Fig. 10. 

There is a weak dependence (if any) of the signal rate on jet core net 
charge, while the inner hadronic bremsstrahlung is predicted to 
grow considerably with it (this prediction 
follows from the coherent nature of the standard hadronic bremsstrahlung
and makes this variable rather interesting from the point of view of 
distinguishing different models of the anomalous soft photon production 
considered in Sect. 8). 

\vskip 0.4cm
{\bf Table 8.} The dependence of direct soft photon rates on the jet core
net charge. 
\begin{center}
\begin{tabular}{ |c c c c c| }
\hline
&&&&\\
$~~Q_{net}~~$&$~~~~<Q_{net}>~~$&$<Q_{net}^{gen}>$&~~~~~~~~~~~~Signal,~~~~~~~~~~~~~&~~~~~Bremsstrahlung,~~~~~~\\
             &                 &                 &     $10^{-3} \gamma$/jet       &  $10^{-3} \gamma$/jet    \\
&&&&\\
\hline
&&&&\\
  0  &  0  &$ 0.19$ & $ ~81\pm ~7\pm 15$ & $ 18.1\pm 0.1\pm 1.2$\\
  1  &  1  &$ 0.93$ & $ ~87\pm ~6\pm 15$ & $ 22.0\pm 0.1\pm 1.4$\\
  2  &  2  &$ 1.70$ & $ ~98\pm 12\pm 20$ & $ 29.5\pm 0.1\pm 1.9$\\
 3,4 & 3.09&$ 2.56$ & $ 106\pm 29\pm 15$ & $ 35.8\pm 0.3\pm 2.3$\\
&&&&\\
\hline
\end{tabular}
\end{center}

\vskip 0.2cm
The linear fit of the bremsstrahlung points results in the slope of the
bremsstrahlung dependence on the jet core net charge to be $(4.66 \pm 0.04)
\times 10^{-3} \gamma$/jet (see solid line in the left upper panel of Fig. 10);
an analogous fit of the signal points (the dashed line in the same panel) 
gives the value of $(7.6 \pm 5.4) \times 10^{-3} \gamma$/jet for the slope. 
Had the signal the same behaviour against this variable as the bremsstrahlung 
has (scaled simply by a factor of 4), a slope of 
$18.6 \times 10^{-3} \gamma$/jet would be expected.
Thus, there is a tendency for the signal dependence on the jet core 
net charge to deviate from the bremsstrahlung behaviour. However, the deviation 
is not significant (about two standard deviations) and  
does not allow a conclusion about an essential difference in the dependences 
of the signal and the bremsstrahlung rates on this variable to be drawn.
Nevertheless, a stronger variation (proportional to the net charge squared) 
which can be assumed for the signal in collective models of the radiation 
(considered below, Sect. 8.2) can be restricted. An upper limit of 27\%
of the signal for the quadratic component of the jet core net charge 
dependence was obtained at 95\% CL. This upper limit was calculated by 
adding a quadratic term to the fit of the dependence, which used a (varied)
constant term together with the fixed bremsstrahlung contribution, 
and increasing the quadratic term yield from zero until the total 
$\chi^2$ increases by 3.84, the 95\% confidence level for the fit with a 
single degree of freedom.  

The dependence of the direct soft photon production rate on the jet core
charged multiplicity has apparently a non-trivial behaviour, decreasing with 
the core $n_{ch}$ increase, and getting closer to the bremsstrahlung 
predictions at higher core $n_{ch}$. This behaviour clearly differs from 
the signal dependence on the $N_{ch}$ multiplicity presented in Table~5 
(such a comparison seems to be most suitable since the cuts on the lower 
charged particle momenta are identical in both cases). 
Nevertheless, there is a significant difference between the two 
selections: in the latter case the selection of charged particles was done 
with the neutral multiplicity being kept fixed at a certain value, while in 
the former case  it was allowed to vary freely. In particular, the averaged 
$N_{neu}$ multiplicity decreases from the value of 2.6 (with the r.m.s. of 1.2)
in the bin with the core $n_{ch} = 0$ to the value of 1.7 (with the r.m.s. 
of 1.1) in the last $n_{ch}$ bin. 
This anti-correlation, induced mainly by the $p_{jet} \geq$ 20 GeV/$c$ cut
(roughly speaking, the smaller is the core $n_{ch}$, the larger 
should be $N_{neu}$ in order to satisfy this cut), 
can be seen in Fig. 1i, where the plot of $N_{neu}$ vs core $n_{ch}$ 
is given. Since the $N_{neu}$ multiplicity appears to be a variable 
which governs the soft photon production in hadronic decays of the $Z^0$, 
this anti-correlation may be responsible for the reduction of the photon rates 
with increasing core $n_{ch}$. However it is difficult to make a final
conclusion on this behaviour until a theoretical description of the
observed anomalous soft photon production will become available. 

\vskip 0.4cm
{\bf Table 9.} The dependence of direct soft photon rates on the jet core
charged multiplicity. 
\begin{center}
\begin{tabular}{ |c c c c c| }
\hline
&&&&\\
~Core $n_{ch}$&~$<$Core $n_{ch}>$&$<$Core $n_{ch}^{gen}>$&~~~~~~~~~~Signal,~~~~~~~~~~~&~~~Bremsstrahlung,~~~~\\
             &                 &                 &     $10^{-3} \gamma$/jet       &  $10^{-3} \gamma$/jet    \\
&&&&\\
\hline
&&&&\\
  0  &  0  & $ 0.13$ & $ 119\pm 12\pm 15$ & $ 16.6\pm 0.1\pm 1.1$ \\
  1  &  1  & $ 1.07$ & $ 112\pm ~8\pm 16$ & $ 22.5\pm 0.1\pm 1.4$ \\
  2  &  2  & $ 2.00$ & $ ~86\pm ~7\pm 16$ & $ 23.8\pm 0.1\pm 1.5$ \\
3$-$5& 3.40& $ 3.23$ & $ ~58\pm ~7\pm 16$ & $ 25.5\pm 0.1\pm 1.6$ \\
&&&&\\
\hline
\end{tabular}
\end{center}
\vskip 0.2cm

\section{ Discussion of the results}
\subsection{General remarks}
What is the source of the direct soft photon signal in hadronic decays of 
the $Z^0$, which exceeds the level of hadronic bremsstrahlung predictions
by a factor of four? Certainly, 25\% of the signal can be attributed to the
bremsstrahlung itself \footnote{ 
It is interesting to note that the subtraction of the bremsstrahlung 
predictions from the signal points measured vs $N_{neu}$ variable (Fig.~6)
makes the resulting distribution (not shown) quite linear, with the fit line 
passing very closely to the origin of the coordinate frame. 
A similar exercise with the signal distribution vs $N_{par}$ (Fig.~7)
improves the $\chi^2$ value of the linear fit mentioned in Sect.~7.4.}.  
Can the rest of the signal be explained by 
an imperfectness of the standard event generators, used in the analysis,
which leads to a huge underestimation of the production of soft photons 
(and may be soft gluons also) in the fragmentation process, or, at least, 
by an imperfectness of the photon implementations \cite{lundgam,arigam,hergam} 
in the generators? In principle, such a  possibility is not excluded. 
However, it looks quite improbable \cite{torn,torn2,boa}, 
unless new physical effects will be introduced to the generator algorithms. 
In this section we shall review in brief the general features of 
theoretical models proposed for the explanation
of the soft photon excess in reactions of multiple hadron production,
and consider their compatibility with the signal behaviour reported in
this work.

The prominent difference of this behaviour from the bremsstrahlung one,
seen in Figs.~6,7, demonstrates that the direct soft photon production in 
hadronic decays of the $Z^0$ depends not only on the charged hadrons produced, 
as it would be for the inner hadronic bremsstrahlung, 
but certainly on the neutral hadrons too. 
Since the direct coupling of photons to neutral particles (e.g. via 
magnetic moment) is quite weak, this means that the excess photons under 
study are coupled either to the individual quarks and/or quark-antiquark
pairs constituting a parton shower, 
%the constituents converting later into final hadrons including the neutral ones, 
or via some collective effect (for example, one of those mentioned in 
\cite{drem}) to a jet as a whole.
 
The first assumption may enter in an apparent conflict with the expected 
damping of the soft photon radiation due to coherent effects known as the
Landau$-$Pomeranchuk$-$ Migdal suppression (LPM effect) \cite{lan,migd,feinb}, 
which in the given case would be due to destructive interference 
between successive photon emitters. 
However, this remark is valid only when the interference between
radiation sources is strong and destructive. In several models 
aiming at an explanation of the anomalous soft photon effect, 
the possibility of interference is discarded or ignored altogether.
In the Van Hove and Lichard model of the cold quark-gluon plasma 
as the source of the soft photons \cite{lich,vanhov,lichard}, the 
photon rate is proportional to the (incoherent) sum of cross sections of the
photon production in head-on collisions of partons, mainly in the processes
of annihilation $(q \overline{q} \rightarrow g \gamma)$ and Compton scattering 
$(q g \rightarrow q \gamma)$ (note however a critical remark to this approach 
with a reference to the LPM effect given in the paper \cite{quack}). 
Also the model \cite{erfi}, based on the Unruh$-$Davies effect (a purely
quantum-mechanical phenomenon which promotes zero-point electromagnetic field 
fluctuations to the level of real quanta \cite{milonni} and leads to the 
thermal radiation from charged particles undergoing acceleration in addition 
to the bremsstrahlung), assumes an incoherent sum of the radiation intensities 
from different quarks. Nachtmann's model of the anomalous soft photons as 
a synchrotron radiation off quarks \cite{nacht,nacht2,nacht3} 
in the stochastic QCD vacuum \cite{dosh} 
also adds the contributions of synchrotron photons from different
partons incoherently. This effectively means that the contribution of each 
quark to the radiation intensity must be proportional to {\em the quark
charge squared}. 

Turning to the models exploiting collective behaviour of radiation sources
(let us call them collective models for brevity), it is interesting to 
note that in Barshay's model of a transient new coherent condition of 
matter \cite{barsh1,barsh2,barsh3,barsh4}, 
proposed for the explanation of the anomalous 
soft photon production in hadronic beam experiments \cite{wa27,na22,wa83},   
the soft photon radiation enhancement appears also due to an explicitly 
non-linear feature of the model, but the radiation itself is coherent and 
the enhancement occurs due to a $constructive$ interference of radiation 
sources \cite{barsh1}.

\subsection{ Collective models of the radiation}
By definition, the collective models assume the presence of some kind of
a medium, or an ensemble of particles (in the case relevant to this study,
it could be a parton shower containing a big number of constituents).
The radiation appearing in the collective models has
notably coherent nature since the collective modes of excitation of the 
medium leading to the radiation are based on the correlations between the  
radiation sources. Therefore the collective models of the radiation pertain 
to coherent models. A classical example is the transition radiation 
induced by a charged particle traversing a boundary between two media with 
different electric polarizability \footnote{An example relevant to the strong 
interactions (in addition to the already mentioned \cite{barsh1}) can be found 
in \cite{drem2} where the model of the coherent hadron production
analogous to Cherenkov radiation is suggested.}. In this case the emerging 
radiation can be considered as a coherent sum of fields emitted by those parts 
of the media (polarized by the particle traversing it)
which are adjacent to the particle trajectory \cite{jack}. In the case of 
the anomalous soft photon production in reaction (1) some combination of 
the charged jet constituents (whatever they are, quarks or the final hadrons)
would be the basic source of the (coherent) radiation
(note, the hadronization time available, namely
100$-$200 fm/$c$ in lab for jets of 45 GeV, see \cite{ddt3}, is big enough 
to allow the formation of soft photons with transverse momenta of 
20$-$80 MeV/$c$, which constitute a bulk of the signal, see \cite{aspdel}, 
thus making the coherent approach to the observations reported in this work 
reasonable). 

In this case the production rate of the anomalous photons should depend on
the collective jet characteristics, jet net charge and mass.
No such dependences were found in the data, as demonstrated by the results 
described in Sects. 7.6 and 7.7. In particular, the quadratic component
of the net charge dependence which can be assumed from Barshay's model
\cite{barsh1} was found to contribute less than 27\% to the signal 
(at 95\% CL).  

Thus the excess photons are unlikely to be produced 
via some collective effects in jets. 

\subsection{ Incoherent models of the radiation}
In these models the production rate of the anomalous soft photons is predicted 
to be proportional to the sum of the charges squared of quarks constituting 
the parton shower. Assuming further the proportionality of the number of 
these quarks to the total jet particle multiplicity, a linear dependence 
between the soft photon rate and the mentioned multiplicity can be predicted. 

The observed dependence of the soft photon production rate on the total jet 
particle multiplicity (Fig. 7) agrees well with this hypothesis. 
A linear fit with zero offset to the coordinate 
system origin displayed in Fig. 7 describes well the experimental points.

However the assumption of the soft photon rate being simply proportional to 
the sum of the quark charges squared is unlikely to be reconciled with the
prominently different dependences of the rates on the jet charged and
neutral particle multiplicities derived in Sect. 7.5. This is a real problem
for incoherent models. 

\subsection{Modification of the incoherent approach}
The difference in the dependences of the photon production rates on the jet 
charged and neutral particle multiplicities can be interpreted more easily 
in the frame of a $q \overline{q}$ dipole model of the radiation, 
the dipoles being formed in a parton shower in the fragmentation 
process. The mean electromagnetic radiation strength of a 
$q \overline{q}$ dipole is expected to be by an order of magnitude higher for 
the neutral $q \overline{q}$ pair than that for the charged one.

This expectation follows from the classical (and non-relativistic) 
consideration of the electric dipole moment of two quarks (the consideration 
of the dipole moments of $qqq$ and $\overline{q} \overline{q} \overline{q}$ 
systems is omitted due to their small admixture in a jet). 
This moment is 
\begin{equation}
\vec{d} = \sum_{i=1}^2 q_i \vec{r}_i~,
\end{equation}
where $q_i$ is the electric charge of the quark $i$, and $\vec{r}_i$ is its
radius-vector pointing to the quark from the origin of the comoving coordinate 
system, which can be fixed at the c.m. position of the quark pair (assuming 
both quark masses to be equal, the origin can be placed at half the distance
between the quarks). The straightforward calculations of dipole moments 
using this formula show that the neutral dipoles, consisting of opposite 
quark charges $\pm 1/3$ or $\pm 2/3$ possess a dipole moment  which is 
higher by a factor of 2 or 4, respectively,
as compared to the charged dipoles consisting of the quark charges
$+1/3, +2/3$ or $-1/3, -2/3$. For the averaged dipole moments squared 
the difference (i.e. the difference in the dipole radiation strength) 
reaches a factor of 10 (note, this estimation of the enhancement factor 
for neutral $q \overline{q}$ dipoles has to be considered as approximate 
as being obtained with formula (3) under the aforementioned assumptions).
 
Decay products of narrow resonances and short-lived unstable particles 
can decrease this contrast when relating the photon rate to the final 
particle multiplicities. 
%(since a certain part of their quark constituents 
%do not participate in the active phase of the quark-gluon jet development). 
Nevertheless, the dependence of the photon production rate on the jet total 
particle multiplicity should remain basically linear, including linear 
components (corresponding to the radiation from neutral and charged 
$q \overline{q}$ pairs), though these components should have different weights. 
Then the following general pattern for the source of  
anomalous soft photon production emerging from the above considerations 
can be suggested. It looks as if $q \overline{q}$ pairs consisting of quarks 
kicked out of the QCD vacuum during the fragmentation process produce 
extra photons $incoherently$ with other $q \overline{q}$ pairs of the jet,
while some coherence inside the pairs (considered as radiating dipoles) 
takes place. 

The pairs can consist of $q \overline{q}$ kicked out of the vacuum
in space-like separated regions, as in the LUND string model
\cite{lundstring} (then some enhancement mechanism is required to explain
the strength of the anomalous soft photon signal, as noticed in \cite{boa}),
or they can appear as closed quark-antiquark loops, as in the model
\cite{simonov,simonov2,simonov3,simonov4,simonov5}, 
which is based on nonperturbative QCD methods applied to
the large size systems and contains a strong enhancement mechanism,
naturally appearing in this approach. The model was primarily
developed for the description of the pion emission by closed
$q \overline{q}$ loops of light quarks inside heavy quarkonia,
but it can be applied also for an analogous description of the soft photon
radiation deep inside jets, which would be a photon source,
additional to the bremsstrahlung radiation from the final state
hadrons. Preliminary estimations of the soft photon intensity done within
this approach look promising \cite{simclas}, and the development of the
photon application of the model is in progress.

However currently the details of the radiation mechanism still remain obscure, 
and a quantitative description of the process by any model  
is still lacking.  

\section{ Conclusion}
An analysis of the direct soft photon production rate as a function
of the parent jet characteristics is presented. It contains a study of 
the dependences of the photon production rates on:
 a) jet momenta; 
 b) jet charged particle multiplicity; 
 c) jet neutral particle multiplicity; 
 d) jet total particle multiplicity; 
 e) jet mass; 
% f) jet Lorentz factor;
 f) jet hardness variable; 
 g) jet core net charge; 
 h) jet core charged multiplicity. 

Apart from the overall excess factor of about four,
a good agreement of the direct soft photon behaviour as compared to that 
of the inner hadronic bremsstrahlung predictions is found for the jet momenta, 
mass and hardness, and a satisfactory agreement for the jet charged 
multiplicity and the jet core net charge. As to the jet neutral and total 
multiplicities, as well as for the jet core charged multiplicity,
a prominent difference of the observed soft photon signal 
from the bremsstrahlung-like behaviour is observed. 
The data especially show that the soft photon production is governed by
the multiplicity of neutral hadrons. This, and the linear dependence of 
the photon rate on the jet total particle multiplicity 
can be interpreted as a proportionality of the anomalous soft photon radiation 
to the total number of quark-antiquark pairs produced in the fragmentation 
process, with the neutral pairs being more effectively radiating than the 
charged ones. These findings suggest that the anomalous soft photons may 
shed light on the formation of the primary hadrons and thereby the quark 
confinement.   

%         Modified on 04-06-1999 by dimartino
%-------------------------------------------------------------------
\subsection*{Acknowledgements}
\vskip 3 mm
We thank Profs. K.G.~Boreskov, F.S.~Dzheparov, B.~French, A.A.~Grigoryan,
A.B.~Kaidalov, O.V.~Kancheli, F.Krauss, A.M.~Kunin, W.~Ochs, L.B.~Okun,
R.M.~Shahoyan, Yu.A.~Simonov, T.~Sj\"ostrand, P.~Sonderegger, H.J.~Specht,
Z.~W\c as and C.Y.~Wong for fruitful discussions.

We are greatly indebted to our technical
collaborators, to the members of the CERN-SL Division for the excellent
performance of the LEP collider, and to the funding agencies for their
support in building and operating the DELPHI detector.\\
We acknowledge in particular the support of \\
Austrian Federal Ministry of Education, Science and Culture,
GZ 616.364/2-III/2a/98, \\
FNRS--FWO, Flanders Institute to encourage scientific and technological
research in the industry (IWT) and Belgian Federal Office for Scientific,
Technical and Cultural affairs (OSTC), Belgium, \\
FINEP, CNPq, CAPES, FUJB and FAPERJ, Brazil, \\
%Czech Ministry of Industry and Trade, GA CR 202/99/1362,\\
%Ministry of Education of the Czech Republic LA134,\\
Ministry of Education of the Czech Republic, project LC527, \\
Academy of Sciences of the Czech Republic, project AV0Z10100502, \\
Commission of the European Communities (DG XII), \\
Direction des Sciences de la Mati$\grave{\mbox{\rm e}}$re, CEA, France, \\
Bundesministerium f$\ddot{\mbox{\rm u}}$r Bildung, Wissenschaft, Forschung
und Technologie, Germany,\\
General Secretariat for Research and Technology, Greece, \\
National Science Foundation (NWO) and Foundation for Research on Matter (FOM),
The Netherlands, \\
Norwegian Research Council,  \\
State Committee for Scientific Research, Poland, SPUB-M/CERN/PO3/DZ296/2000,
SPUB-M/CERN/PO3/DZ297/2000, 2P03B 104 19 and 2P03B 69 23(2002-2004),\\
FCT - Funda\c{c}\~ao para a Ci\^encia e Tecnologia, Portugal, \\
Vedecka grantova agentura MS SR, Slovakia, Nr. 95/5195/134, \\
Ministry of Science and Technology of the Republic of Slovenia, \\
CICYT, Spain, AEN99-0950 and AEN99-0761,  \\
The Swedish Research Council,      \\
The Science and Technology Facilities Council, UK, \\
%Particle Physics and Astronomy Research Council, UK, \\
Department of Energy, USA, DE-FG02-01ER41155, \\
EEC RTN contract HPRN-CT-00292-2002. \\

\newpage

\begin{figure}[0]
\begin{center}
\epsfig{file=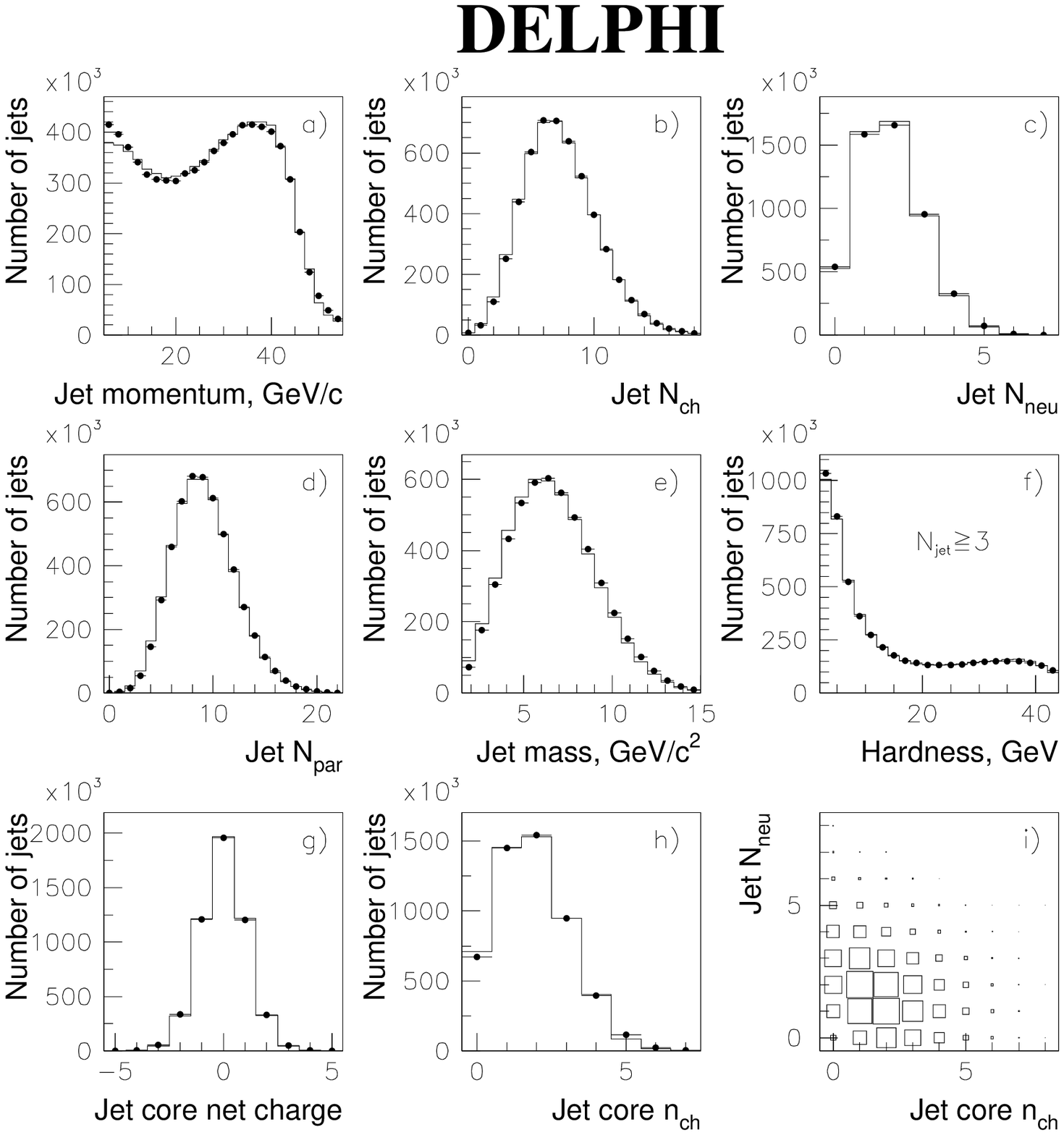,bbllx=50pt,bblly=180pt,bburx=550pt,bbury=570pt,%
width=17cm,angle=0}
\end{center}
\caption{a) to h), the distributions of the variables used in this analysis,
obtained with both, the real data (points), and with the MC (histograms); 
a) jet momentum; b) jet charged multiplicity; c)~jet neutral multiplicity;
d) jet total multiplicity; e) jet mass; f)~hardness variable, $\kappa_J$;
g) jet core net charge; h) jet core charged multiplicity. The panel~i) 
shows the correlation plot of the jet $N_{neu}$  vs the jet core $n_{ch}$.}
\end{figure}

\begin{figure}[1]
\begin{center}
\epsfig{file=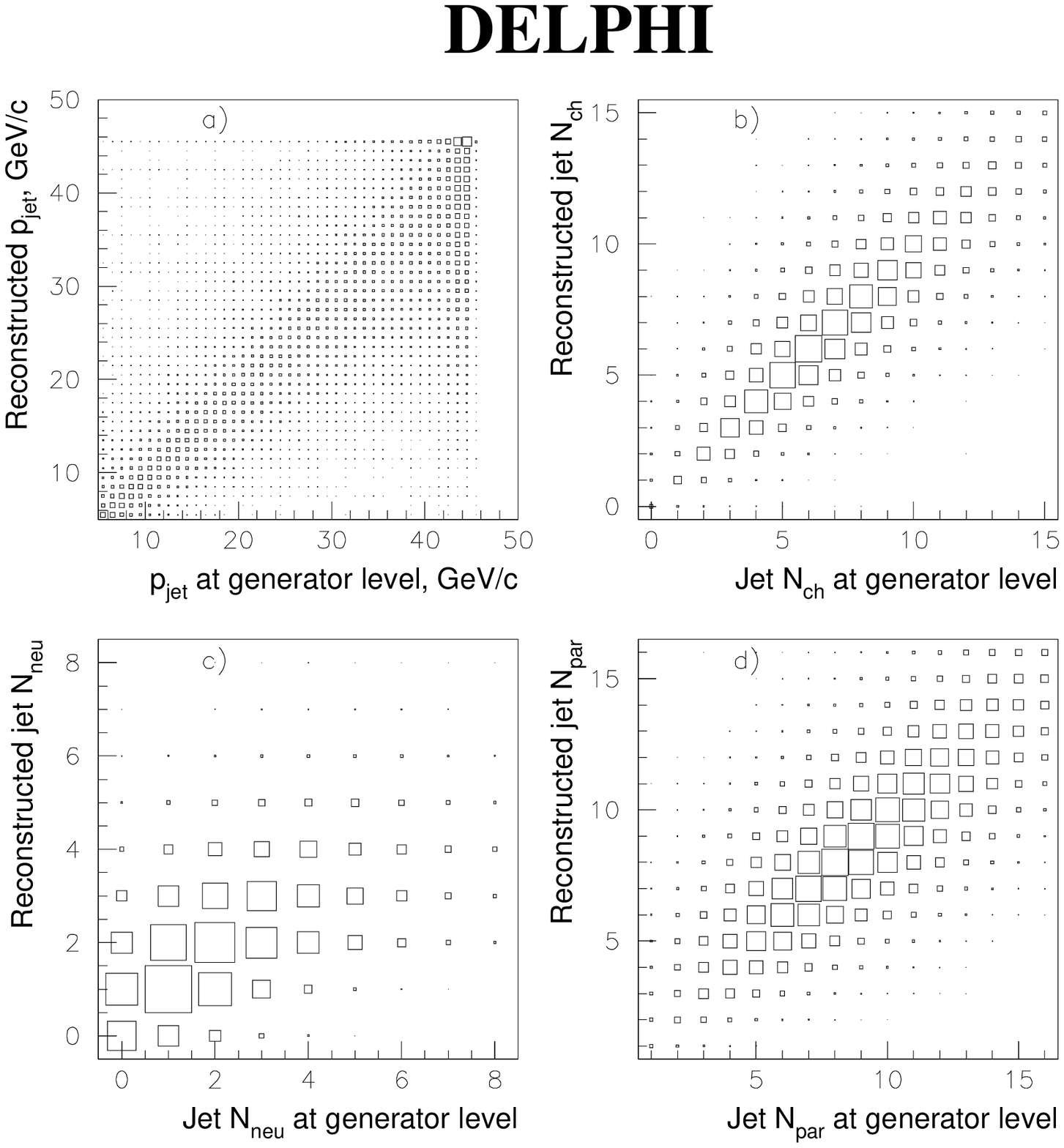,bbllx=50pt,bblly=180pt,bburx=550pt,bbury=570pt,%
width=17cm,angle=0}
\end{center}
\caption{a) Reconstructed jet momentum $p_{jet}$ vs jet momentum at the 
generator level;
b) reconstructed jet $N_{ch}$ multiplicity vs jet charged multiplicity
at the generator level;
~~c) reconstructed jet $N_{neu}$ multiplicity vs jet neutral multiplicity
at the generator level;
~~d) reconstructed jet $N_{par}$ multiplicity vs jet particle multiplicity
at the generator level.}
\end{figure}
\begin{figure}[2]
\begin{center}
\epsfig{file=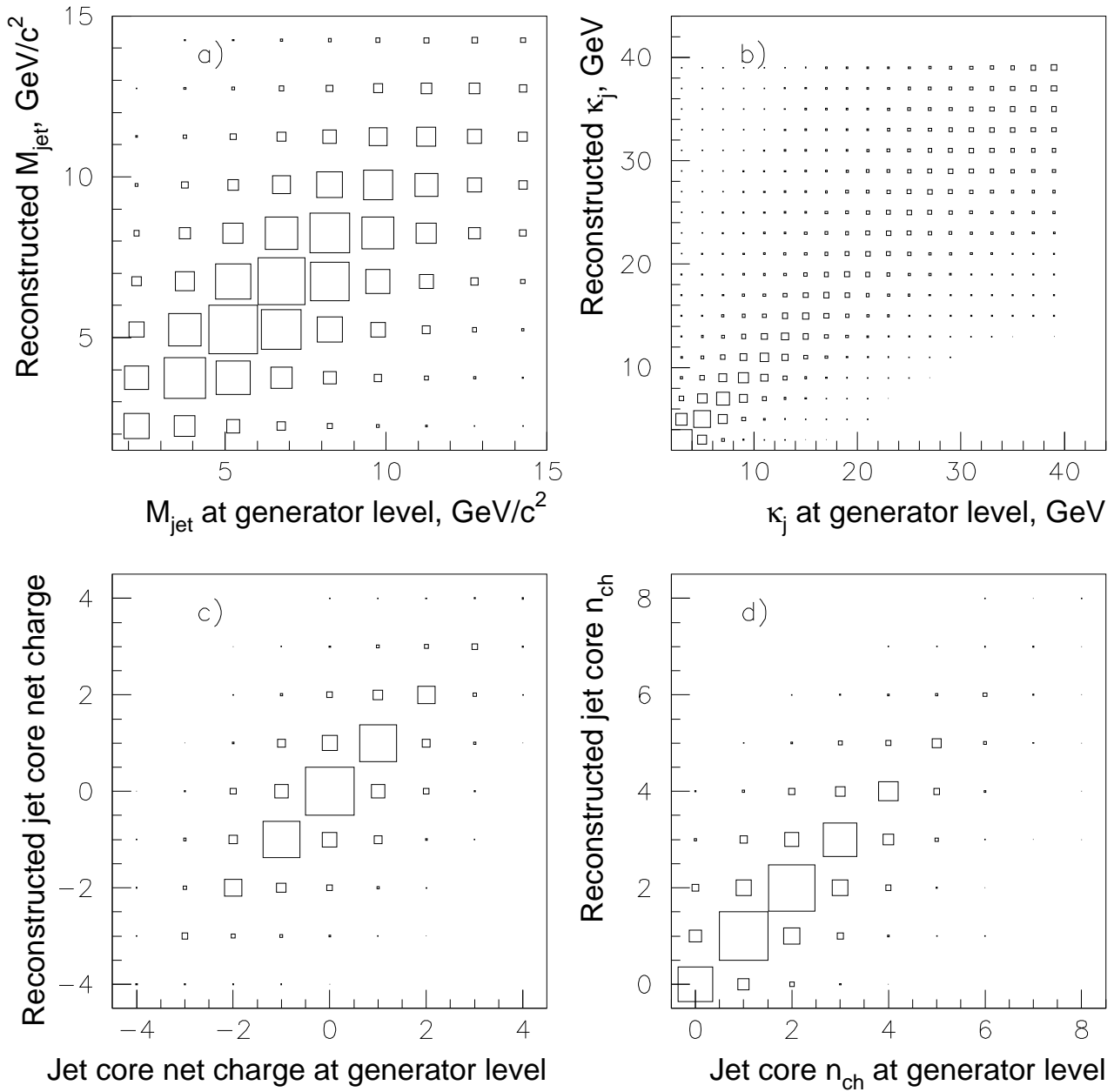,bbllx=50pt,bblly=180pt,bburx=550pt,bbury=570pt,%
width=17cm,angle=0}
\end{center}
\caption{a) Reconstructed jet mass $M_{jet}$ vs jet mass at the generator level;
%b) reconstructed jet Lorentz factor vs jet Lorentz factor at the 
b) reconstructed jet variable $\kappa_J$ vs $\kappa_J$ at the generator level;
c) reconstructed jet core net charge vs jet core net charge at 
the generator level;
d) reconstructed jet core $n_{ch}$ vs jet core $n_{ch}$ at the generator level.}
\end{figure}
                                                                                
\begin{figure}[3]
\begin{center}
\epsfig{file=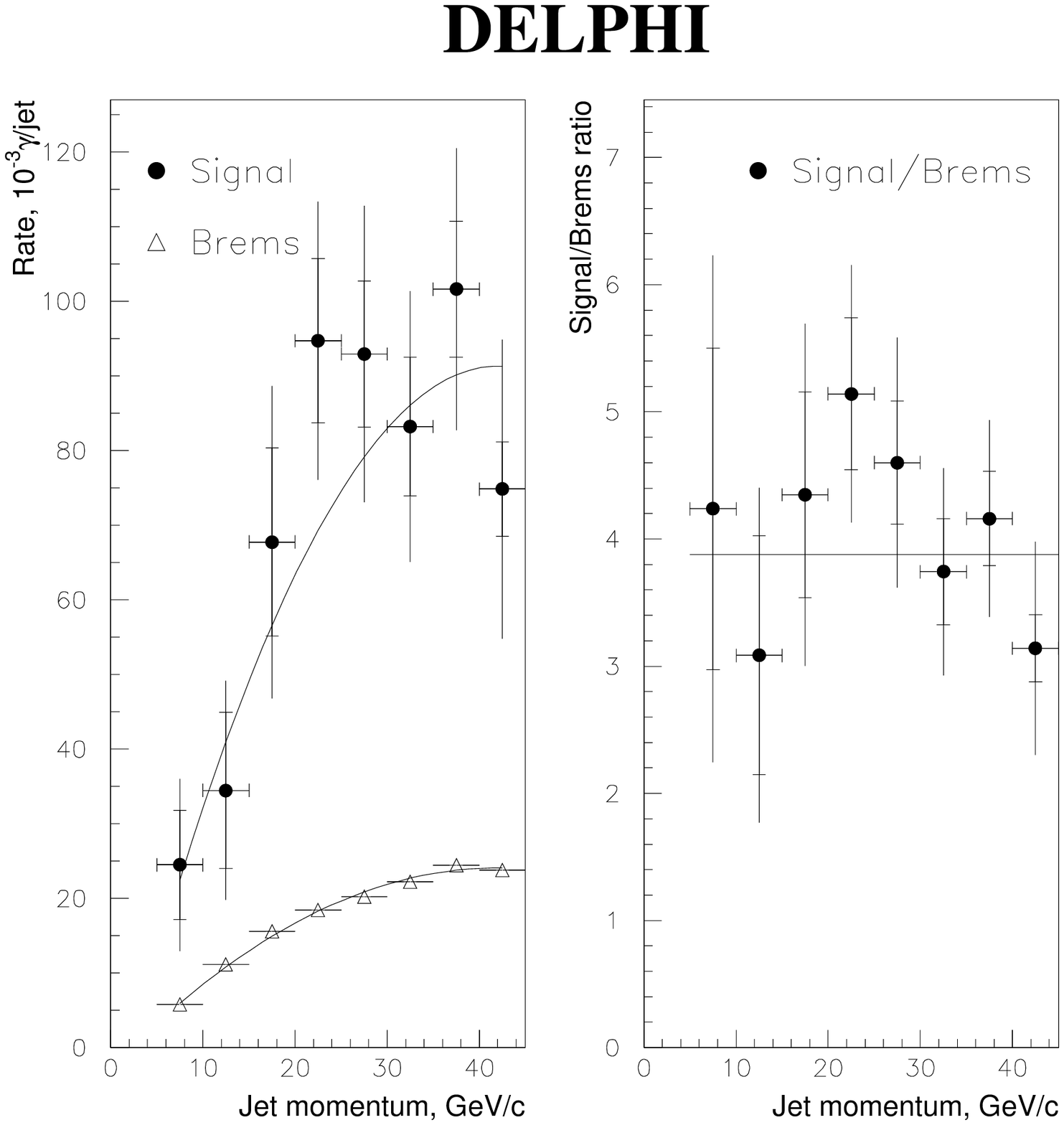,bbllx=50pt,bblly=180pt,bburx=550pt,bbury=570pt,%
width=17cm,angle=0}
\end{center}
\caption{Dependence of the direct soft photon production on the jet momentum. 
Left panel: signal and predicted inner bremsstrahlung rates as a function 
of jet momentum. Right panel: ratios of the signal rates to those of the inner 
bremsstrahlung. The curves in the left panel are 2nd order polynomial fits 
produced to guide the eye; the bremsstrahlung points were fitted first, 
and then the bremsstrahlung curve was scaled by a factor of 4 giving a good 
approximation to the signal points. The inner vertical bars represent the
statistical errors, while the whole vertical bars give the statistical and
systematic errors combined in quadrature. 
The horizontal line in the right panel represents the statistical 
average over the signal-to-bremsstrahlung ratios.}
\end{figure}

\begin{figure}[4]
\begin{center}
\epsfig{file=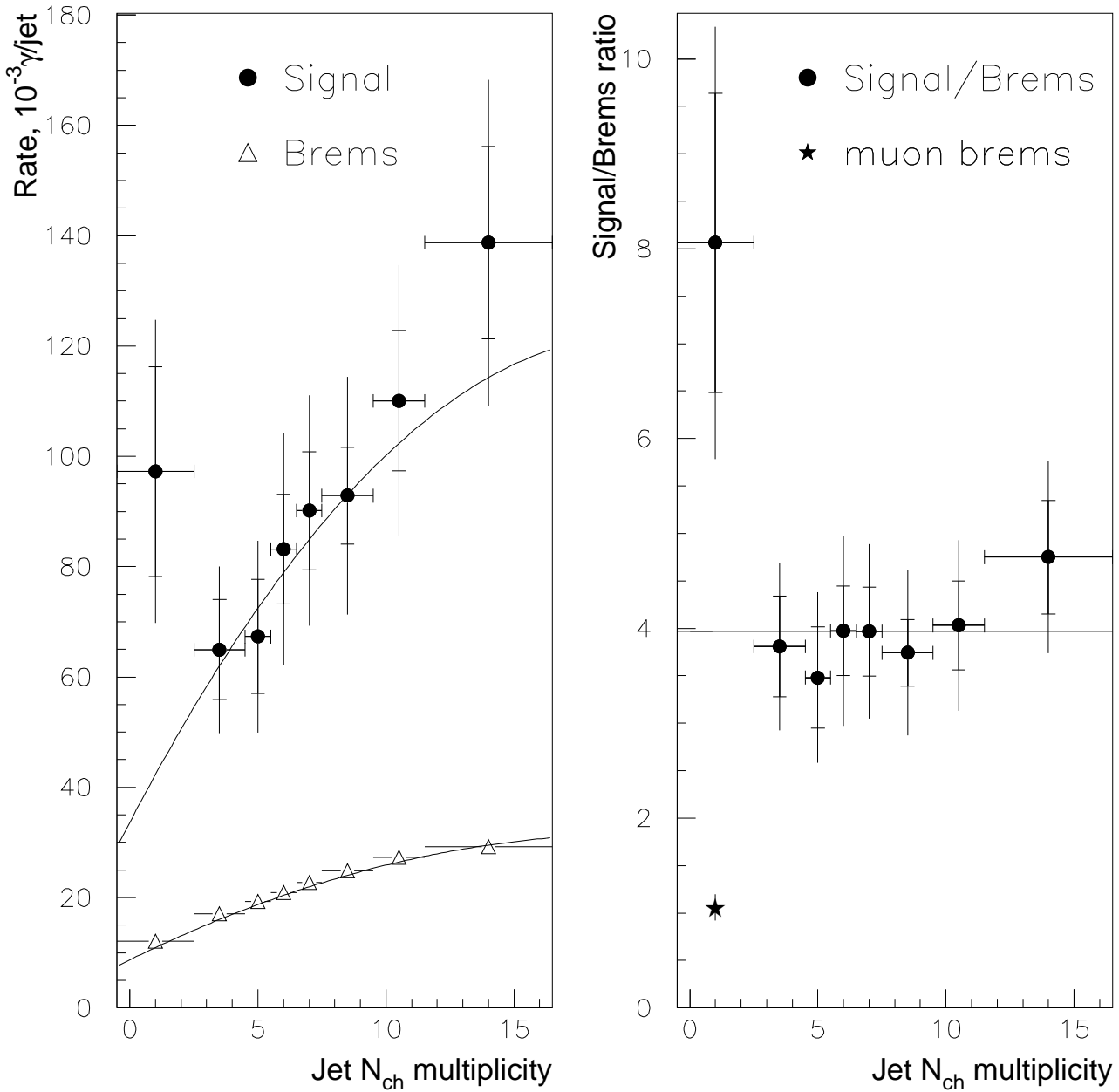,bbllx=50pt,bblly=180pt,bburx=550pt,bbury=570pt,%
width=17cm,angle=0}
\end{center}
\caption{Dependence of the direct soft photon production on the jet charged
multiplicity. Left panel: signal and predicted inner bremsstrahlung rates 
as a function of the jet charged multiplicity.
Right panel: ratios of the signal rates to those of the inner bremsstrahlung.
The curves in the left panel are 2nd order polynomial fits 
%(with restrictions on the constant term, see text, Sect. 7.2) 
produced to guide the eye; the bremsstrahlung points
were fitted first, and then the bremsstrahlung curve was scaled by a factor 
of 4,  which satisfactorily approximates the signal points. 
The inner vertical bars represent the
statistical errors, while the whole vertical bars give the statistical and
systematic errors combined in quadrature. The horizontal 
line in the right panel represents the statistical average over the 
signal-to-bremsstrahlung ratios. The cut $p_{jet} \geq$ 20 GeV/$c$ is applied.}
\end{figure}

\begin{figure}[5]
\begin{center}
\epsfig{file=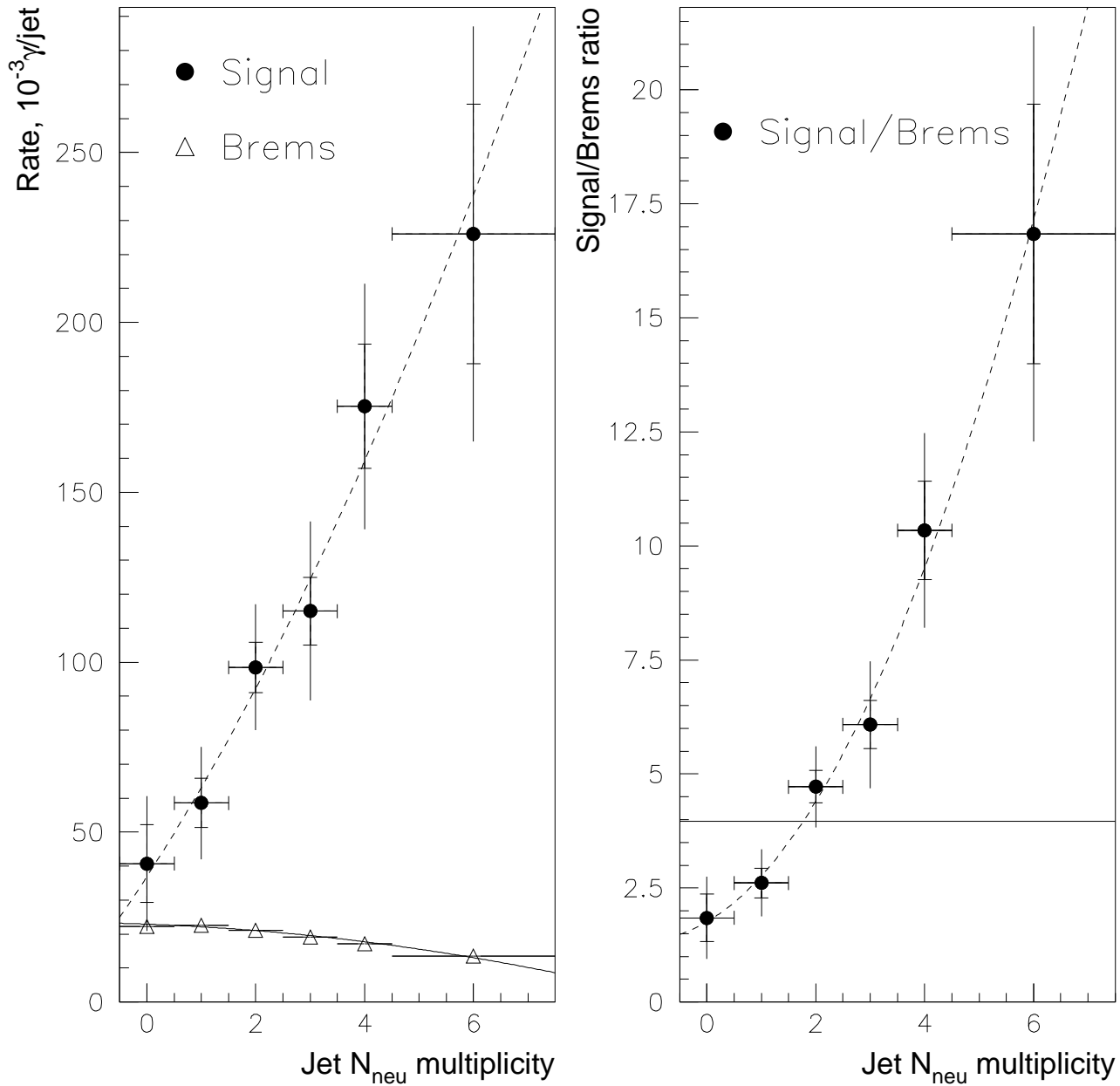,bbllx=50pt,bblly=180pt,bburx=550pt,bbury=570pt,%
width=17cm,angle=0}
\end{center}
\caption{Dependence of the direct soft photon production on the jet neutral
multiplicity. Left panel: signal and predicted inner bremsstrahlung rates 
as a function of the jet neutral multiplicity.
Right panel: ratios of the signal rates to those of the inner bremsstrahlung.
All the curves in the figure are independent 2nd order polynomial fits 
produced to guide the eye. 
The inner vertical bars represent the
statistical errors, while the whole vertical bars give the statistical and
systematic errors combined in quadrature. The horizontal line in the right 
panel represents the statistical average over the signal-to-bremsstrahlung 
ratios. The cut $p_{jet} \geq$ 20 GeV/$c$ is applied.}
\end{figure}

\begin{figure}[6]
\begin{center}
\epsfig{file=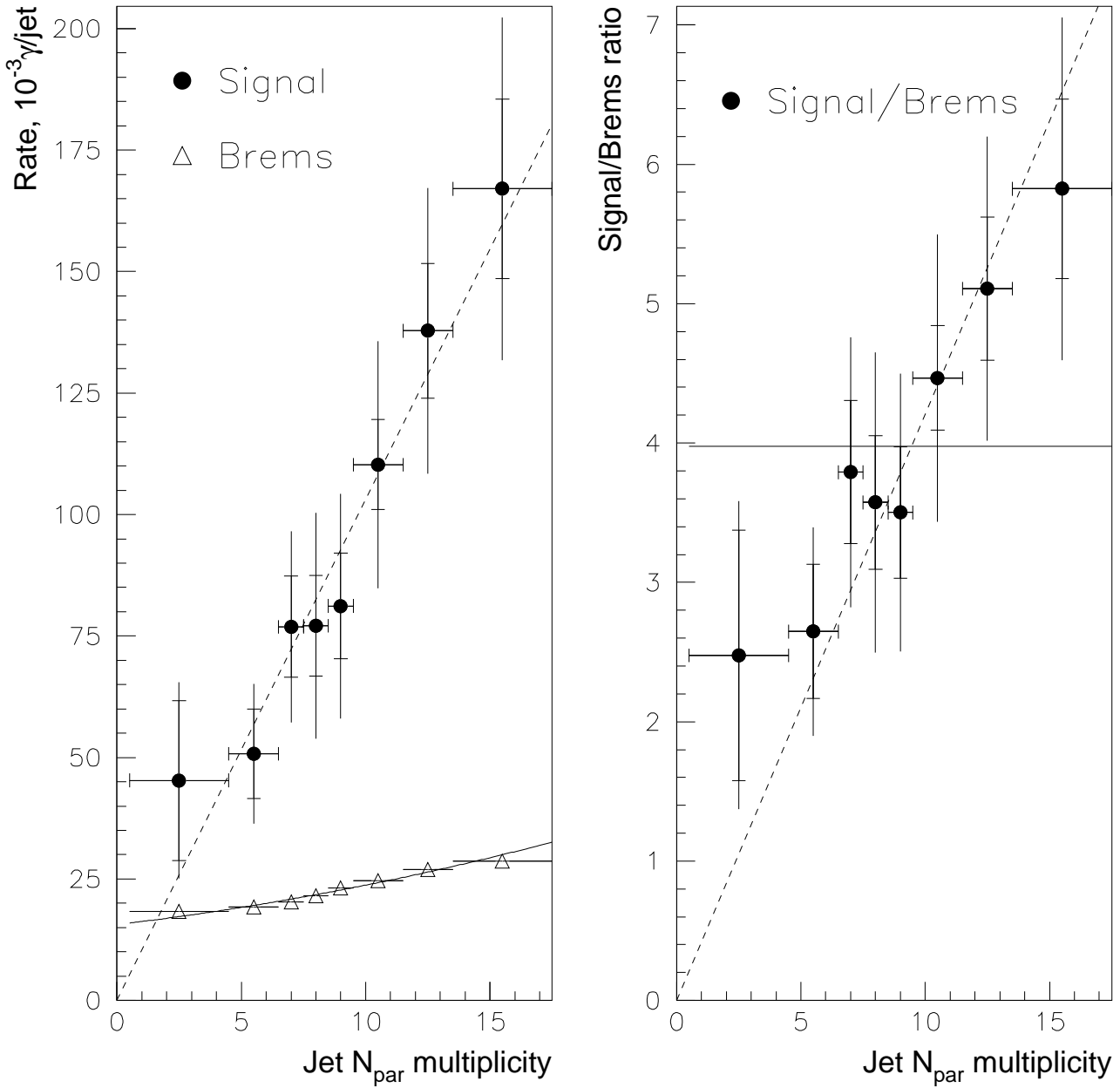,bbllx=50pt,bblly=180pt,bburx=550pt,bbury=570pt,%
width=17cm,angle=0}
\end{center}
\caption{Dependence of the direct soft photon production on the jet total
multiplicity. Left panel: signal and predicted inner bremsstrahlung rates 
as a function of the jet total multiplicity.
Right panel: ratios of the signal rates to those of the inner bremsstrahlung.
The curve through the bremsstrahlung points in the left panel is a 2nd order
polynomial fit produced to guide the eye. The dashed lines in both
panels represent the linear fits of the signal points
with zero offset to the origin of the coordinate system. 
The inner vertical bars represent the
statistical errors, while the whole vertical bars give the statistical and
systematic errors combined in quadrature. The horizontal 
line in the right panel represents the statistical average over the 
signal-to-bremsstrahlung ratios. The cut $p_{jet} \geq$ 20 GeV/$c$ is applied.}
\end{figure}

\begin{figure}[7]
\begin{center}
\epsfig{file=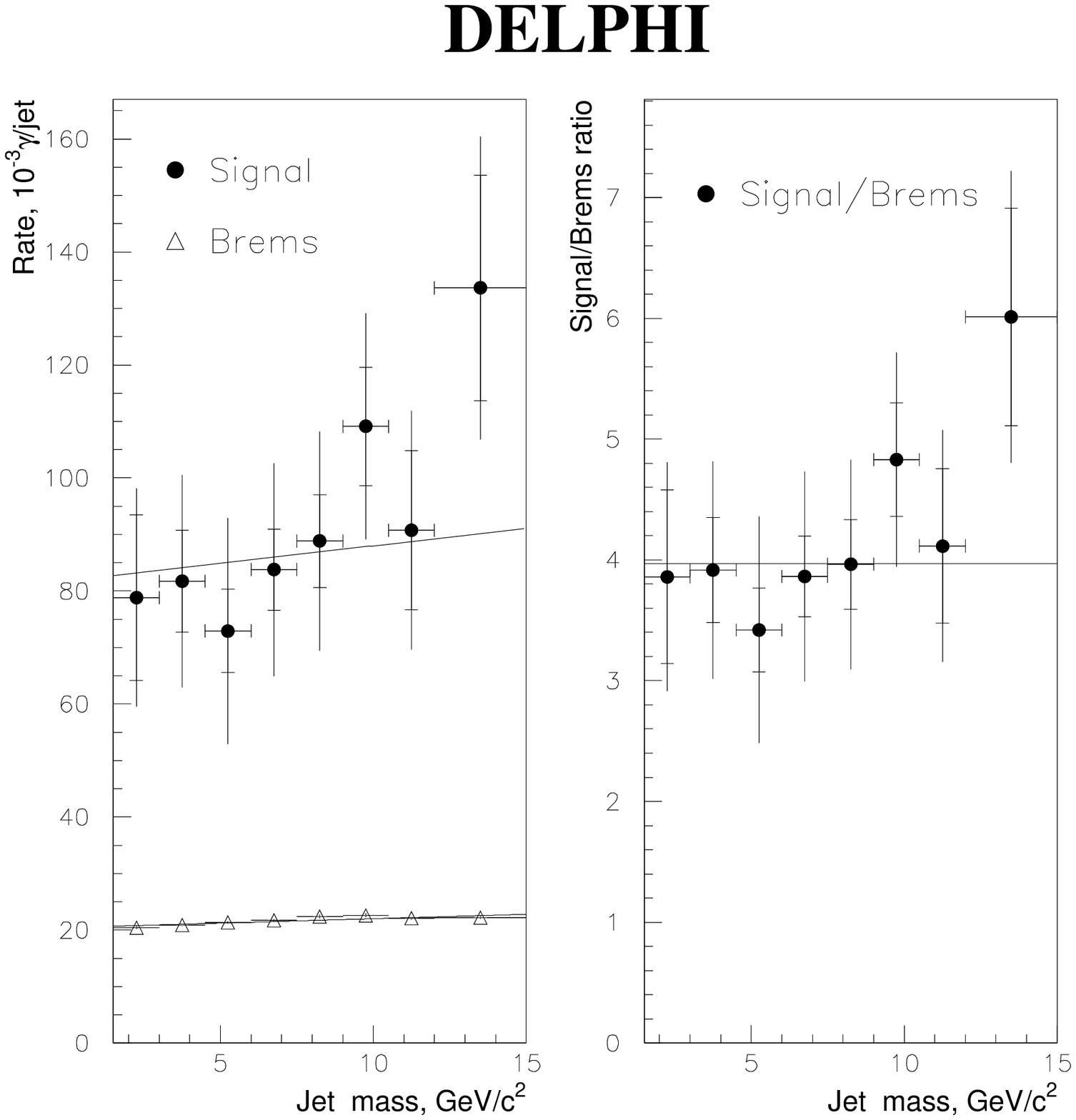,bbllx=50pt,bblly=180pt,bburx=550pt,bbury=570pt,%
width=17cm,angle=0}
\end{center}
\caption{Dependence of the direct soft photon production on the jet mass. 
Left panel: signal and predicted inner bremsstrahlung rates as a function 
of jet mass. Right panel: ratios of the signal rates to those of the inner 
bremsstrahlung. The curves in the left panel are 1st order polynomial fits
produced to guide the eye; the bremsstrahlung points were fitted first,
and then the bremsstrahlung curve was scaled by a factor of 4 giving a good
approximation to the signal points. The inner vertical bars represent the
statistical errors, while the whole vertical bars give the statistical and
systematic errors combined in quadrature. The horizontal line in the right 
panel represents the statistical average over the signal-to-bremsstrahlung 
ratios. The cut $p_{jet} \geq$ 20 GeV/$c$ is applied.}
\end{figure}

\begin{figure}[8]
\begin{center}
\epsfig{file=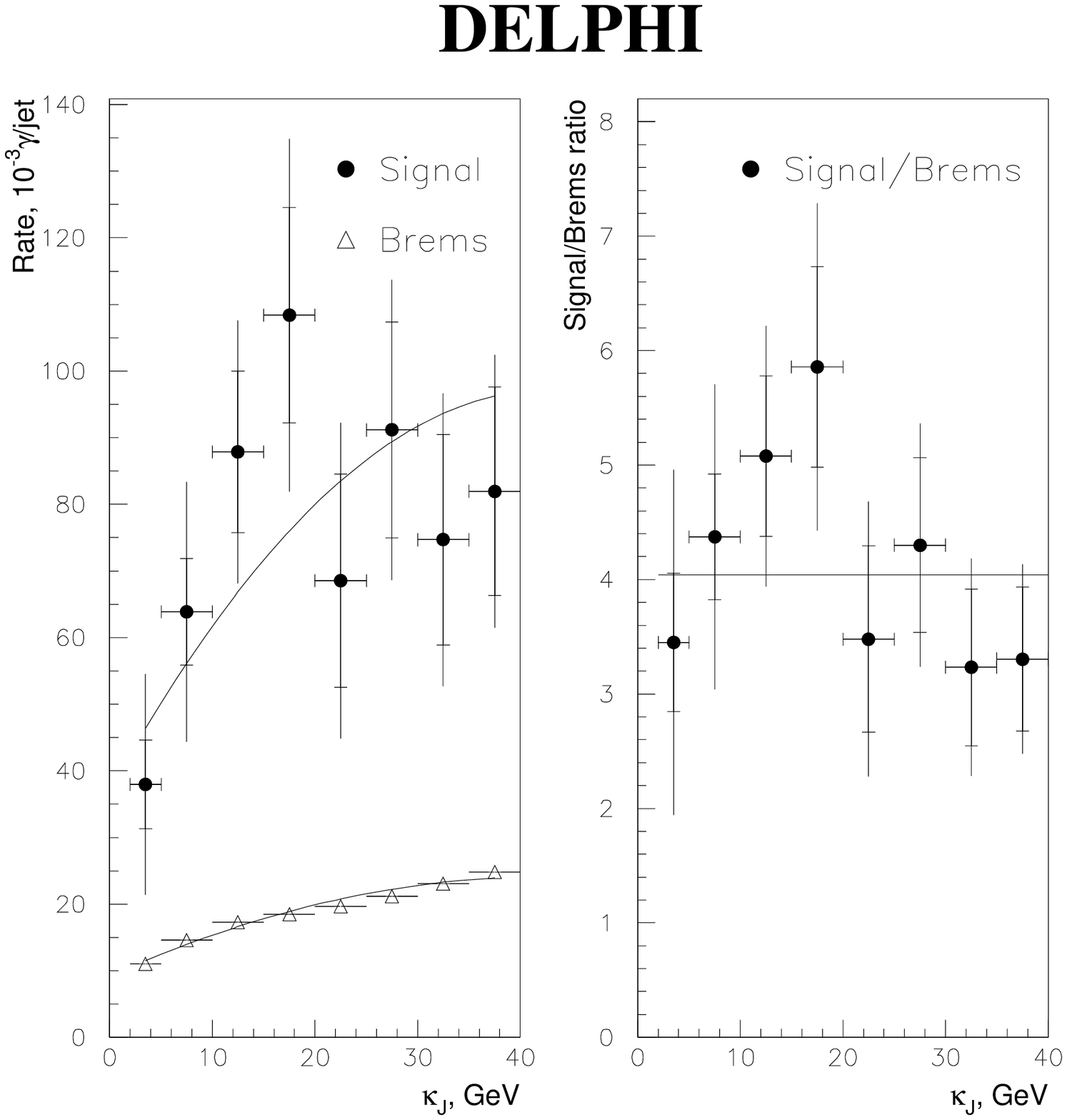,bbllx=50pt,bblly=180pt,bburx=550pt,bbury=570pt,%
width=17cm,angle=0}
\end{center}
\caption{Dependence of the direct soft photon production on the hardness 
variable. Left panel: signal and predicted inner bremsstrahlung rates as a 
function of $\kappa_{j}$. Right panel: ratios of the signal rates to those 
of the inner bremsstrahlung. The curves in the left panel are 2nd order 
polynomial fits produced to guide the eye; the bremsstrahlung points were 
fitted first, and then the bremsstrahlung curve was scaled by a factor of 4 
giving a good approximation to the signal points. 
The inner vertical bars represent the
statistical errors, while the whole vertical bars give the statistical and
systematic errors combined in quadrature. The horizontal line 
in the right panel represents the statistical average over the 
signal-to-bremsstrahlung ratios. The cut $p_{jet} \geq$ 20 GeV/$c$ is $not$
applied.}
\end{figure}

\begin{figure}[9]
\begin{center}
\epsfig{file=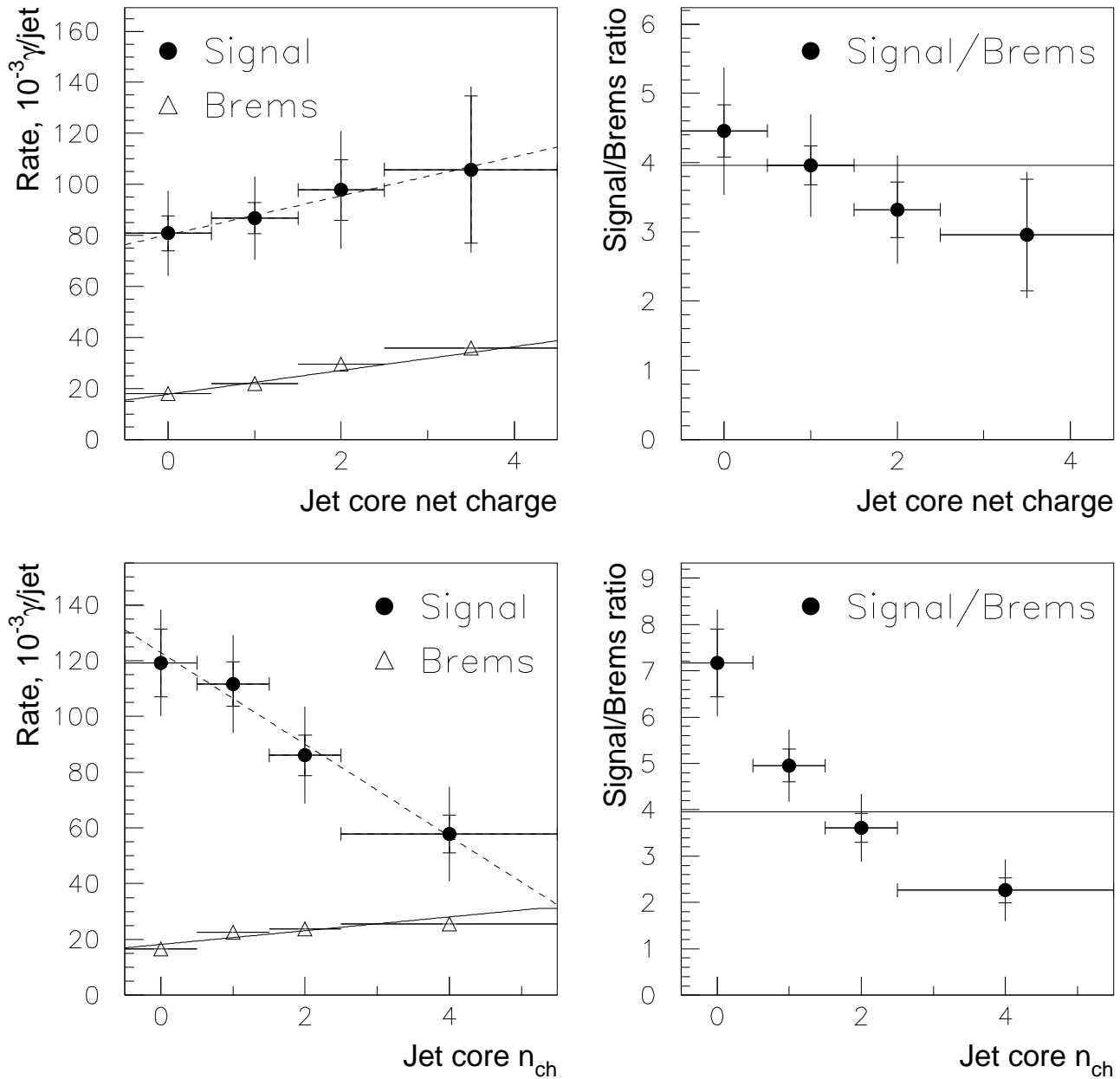,bbllx=50pt,bblly=180pt,bburx=550pt,bbury=570pt,%
width=17cm,angle=0}
\end{center}
\caption{Dependence of the direct soft photon production on the jet core
characteristics. Upper panels: jet core net charge; bottom panels: 
jet core charged multiplicity. Left panels: signal and predicted inner 
bremsstrahlung rates as a function of the jet core characteristics. 
Right panels: ratios of the signal rates to those of the 
inner bremsstrahlung. The straight lines in the left panels are linear fits 
produced to guide the eye: solid line for the bremsstrahlung points and 
the dashed line for the signal. The inner vertical bars represent the
statistical errors, while the whole vertical bars give the statistical and
systematic errors combined in quadrature. The horizontal lines in the right 
panels represent the statistical averages over the signal-to-bremsstrahlung 
ratios. The cut $p_{jet} \geq$ 20 GeV/$c$ is applied.}
\end{figure}
%=========================================================================%

\end{document}